\documentclass[prd,nofootinbib,floats,superscriptaddress,eqsecnum,tightenlines,preprintnumbers,11pt]{revtex4-1}

\usepackage{hyperref}
\usepackage{graphicx}
\usepackage{amsmath,amssymb,amsfonts,amsthm,latexsym,stmaryrd}
\usepackage{marginnote}
\usepackage{color}

\usepackage{dsfont}
\usepackage{enumerate}
\usepackage{float}
\def\be{\begin{equation}}
\def\ee{\end{equation}}
\def\bea{\begin{eqnarray}}
\def\eea{\end{eqnarray}}
\def\beq{\begin{eqnarray}}
\def\eeq{\end{eqnarray}}
\def\bas{\begin{subequations}\begin{eqnarray}}
\def\eas{\end{eqnarray}\end{subequations}}
\def\nn{\nonumber}

\def\lsim{\mathrel{\mathop  {\hbox{\lower0.5ex\hbox{$\sim$}
\kern-0.8em\lower-0.7ex\hbox{$<$}}}}}
\def\gsim{\mathrel{\mathop  {\hbox{\lower0.5ex\hbox{$\sim$}
\kern-0.8em\lower-0.7ex\hbox{$>$}}}}}

\def\be{\begin{equation}}
\def\ee{\end{equation}}
\def\bea{\begin{eqnarray}}
\def\eea{\end{eqnarray}}
\def\beq{\begin{eqnarray}}
\def\eeq{\end{eqnarray}}
\def\bas{\begin{subequations}\begin{eqnarray}}
\def\eas{\end{eqnarray}\end{subequations}}
\def\nn{\nonumber}

\newcommand{\cE}{{\mathcal E}}

\newcommand{\cH}{{\mathcal H}}
\newcommand{\cM}{{\mathcal M}}

\newcommand{\cT}{{\mathcal T}}

\newcommand{\cS}{{\mathcal S}}

\newcommand{\bes}{\begin{eqnarray}}
\newcommand{\ees}{\end{eqnarray}}

\def\nn{\nonumber}

\def\dd{\mathrm{d}}

\numberwithin{equation}{section}

\begin{document}

\preprint{YITP-20-51}

\title{Bouncing compact objects. \\
Part II: Effective theory of a pulsating Planck star}

\author{Jibril Ben Achour}
\affiliation{Center for Gravitational Physics, Yukawa Institute for Theoretical Physics, Kyoto University, 606-8502, Kyoto, Japan}
\author{Jean-Philippe Uzan}
\affiliation{Institut d'Astrophysique de Paris, CNRS UMR 7095, Universit\'e Pierre et Marie Curie - Paris VI, 98 bis Boulevard Arago, 75014 Paris, France \\
           Sorbonne Universit\'es, Institut Lagrange de Paris}

\date{\today}

\begin{abstract}
This article presents an effective quantum extension of the seminal Oppenheimer-Snyder (OS) collapse in which the singularity resolution is modeled using the effective dynamics of the spatially closed loop quantum cosmology. Imposing the minimal junction conditions, namely the  Israel-Darmois conditions, we glue this bouncing LQC geometry to the classical vacuum exterior Schwarzschild geometry across a time-like thin-shell. Consistency of the construction leads to several major deviations from the classical OS collapse model. Firstly, no trapped region can form and the bounce occurs always above, or at most at the Schwarzschild radius. Secondly, the bouncing star discussed here admits an IR  cut-off, additionally to the UV cut-off and corresponds therefore to a pulsating compact object. Thirdly, the scale at which quantum gravity effects become non-negligible is encoded in the ratio between the UV cut-off of the quantum theory and the IR cut-off, which in turn, encodes the minimal energy density $\rho_{\text{min}}$ of the star prior to collapse. This energy density is no more fixed by the mass and maximal radius as in the classical OS model, but is now a free parameter of the model. In the end, while the present model cannot describe a black-to-white hole bounce as initially suggested by the Planck star model, it provides a concrete realization of a pulsating compact object based on LQC techniques. Consistency of the model shows that its regime of applicability is restricted to Planckian relics while macroscopic stellar objects are excluded. This first minimal construction should serve as a platform for further investigations in order to explore the physics of bouncing compact objects within the framework of loop quantum cosmology.
\end{abstract}

\maketitle
\tableofcontents

\section{Introduction} 

The existence of ultra compact objects is now routinely confirmed by the newly born gravitational waves astronomy. While black hole solutions of General Relativity (GR) provide to date the best model to fit the associated data, the existence of singularities in their core prevent them from providing a complete and self consistent UV description for these ultra compact objects. The resolution of these classical singularities can be addressed only within a quantum theory of gravity, in which the fate of the quantum geometry at high energy can be answered non perturbatively. As it turns out, the consequence of singularity resolution might be especially relevant from an observational point of view. Indeed, it is worth keeping in mind that these astronomical ultra compact objects being certainly not described by the stationary solutions of GR all the way down to the Planck scale, the absence of singularity in their core might have interesting observational effects. Since no complete quantum theory of gravity is yet available, the common strategy to capture the description of UV complete compact objects is to resort on effective approaches. 

Over the last five years, important efforts have been devoted in investigating the consequences of singularity resolution within the classical gravitational collapse scenario. Among different proposals, the Planck star model presented in Ref. \cite{Rovelli:2014cta} has suggested new fascinating perspectives regarding the final stage of evaporating compact objects. Motivated by results in loop quantum cosmology, see e.g. Ref. \cite{Ashtekar:2011ni} for a review, it was proposed that, after forming a trapping horizon, a collapsing star might form a Planckian core, and subsequently bounces out due to the repulsive quantum gravity pressure, releasing the information stored in the long-lived trapped region. Two crucial ideas were advocated to model this scenario. First, it was pointed that the energy scale at which quantum gravity effects become non-negligible \textit{is not} dictated by the size of the object w.r.t to the Planck length, i.e. $\ell/ \ell_{\text{Planck}}$,  but by its density w.r.t the Planck density, i.e. $\rho/\rho_{\text{Planck}}$, allowing therefore quantum gravity to kick off much above the Planck length. Moreover, it was suggested that while the bounce can be very short in the proper frame of the star, it appears extremely slow for an observer at infinity due to the huge gravitational time dilatation, allowing the model to be consistent with observational constraints. This elegant mechanism was applied to primordial black holes and phenomenological consequences were investigated in Refs. \cite{Barrau:2014hda, Barrau:2015uca, Barrau:2016fcg, Rovelli:2017zoa, Barrau:2018kyv}, suggesting a new observational window towards quantum gravity. More refined constructions of the dynamical geometry describing such bouncing compact object, and in particular the tunneling between the trapped to the anti-trapped region, were investigated later on, leading to the black hole firework and its further generalizations \cite{Haggard:2014rza, DeLorenzo:2015gtx, Rovelli:2018cbg, Brahma:2018cgr, DAmbrosio:2018wgv}. See also Refs. \cite{Barcelo:2014cla, Barcelo:2015uff, Barcelo:2016hgb} for an alternative construction of a black-to-white hole tunelling, based on a radically different mechanism. Finally, efforts to take into account the evaporation process were also discussed recently in Refs.~\cite{Bianchi:2018mml, Rovelli:2018hba, Rovelli:2018okm, Martin-Dussaud:2019wqc}. See also Refs. \cite{Kawai:2013mda, Baccetti:2018qrp, Ho:2018fwq, Ho:2019pjr, Bolokhov:2018rsa} for recent investigations on the semi-classical description of evaporating and collapsing null shells. 

This body of results suggests a new radically different scenario for the final stage of an evaporating and collapsing object which beg for further developments. In particular, can we construct a consistent model of a dynamical Planck star where the singularity resolution in the interior region is modeled by loop quantum cosmology techniques? Important efforts have been devoted recently to implement such techniques to discussed polymer interior black holes, which correspond to vacuum homogeneous geometries, see e.g. Refs. \cite{Corichi:2015xia, Olmedo:2017lvt, Cortez:2017alh, BenAchour:2018khr, Bojowald:2018xxu, Ashtekar:2018cay, Bodendorfer:2019xbp, Bojowald:2019dry, Bouhmadi-Lopez:2019hpp, Bodendorfer:2019cyv, Bodendorfer:2019nvy, Bodendorfer:2019jay, Bojowald:2015zha, Aruga:2019dwq, BenAchour:2017ivq, Assanioussi:2019twp,Morales-Tecotl:2018ugi} and more recently Refs. \cite{Alesci:2018loi, Alesci:2019pbs} for details\footnote{See also Refs. \cite{Ashtekar:2005qt, Modesto:2006mx, Bohmer:2007wi, Boehmer:2008fz, Modesto:2009ve, Chiou:2012pg, Tibrewala:2012xb, Gambini:2013ooa} for earlier works on this topic.}. In order to develop a Planck star model, one needs to go beyond this framework and consistently include the role of collapsing matter. Previous attempts in this direction were presented in Refs. \cite{Bojowald:2005qw, Tavakoli:2013lga, Tavakoli:2013rna, Bambi:2013caa, Liu:2014kra, These, Joe:2014tca, Gambini:2014qga} as well as in Refs. \cite{BenAchour:2016brs, BenAchour:2017jof, Bojowald:2019fkv} following a different approach. In this work, we adopt a different strategy. Using the quantum extension of the Oppenheimer-Snyder model presented in our companion article\cite{paperUS}, we present a new construction, based on the thin shell formalism, which allows one to discuss matter collapse using LQC techniques, and provides therefore a minimal set up to realize concretely, within the LQC framework, part of the ideas introduced in the initial Planck star model in Ref. \cite{Rovelli:2014cta}.

To that goal, the interior homogenous geometry will be modeled by the loop quantum spatially closed cosmology filled with dust, discussed in Refs. \cite{Corichi:2011pg, Dupuy:2016upu}, while the exterior will be the classical Schwarzschild geometry. The singularity resolution, which manifests through a bounce at some critical energy, requires the existence of a non-vanishing energy-momentum localized on the thin shell joining the exterior and interior geometries. This time-like thin-shell plays a central role in that it encodes part of the quantum effects. In order to find admissible solutions for its surface energy and pressure, a key condition has to be satisfied, which arises from the generalization of the standard mass relation found by Oppenheimer and Snyder \cite{Oppenheimer:1939ue}. It ensures that the constants of motion associated to the exterior and interior geometries properly match during the whole process. Additionnaly, it translates into a constraint on the energy scale at which the bounce, and therefore the quantum gravity effects, becomes non-negligible.

In this regard, the present model allows one to implement the different ideas discussed earlier. It implements the singularity resolution mechanism using LQC techniques, it includes the role of matter collapse which is modeled by a dust fluid, and it describes both the exterior and interior geometry of a UV complete bouncing compact object while ensuring the conservation laws during the whole process. As such, this simple model provides the minimal set up to discuss an effective UV complete gravitational collapse and its internal consistency using the techniques of LQC. Let us emphasize the major outcomes of this construction.

Under the set of hypothesis discussed above, it was shown in Ref. \cite{paperUS} that such idealized bouncing object does not form a trapped region. The compact object forms at best a marginally trapping horizon, but automatically bounces at most at the energy threshold of horizon formation. As such, no black hole (as defined by the existence of a trapped region) is formed. This result provides a major deviation from the classical Oppenheimer-Snyder model where an event horizon and a subsequent singularity are formed as a result of the collapse. Moreover, it directly implies that quantum effects dominate at scales larger than or at the Schwarzschild radius, where the bounce occurs. As emphasized in Ref. \cite{paperUS}, this is a direct consequence of demanding continuity of the induced metric across the time-like thin-shell, as well as working with a vacuum classical Schwarzschild exterior geometry. Hence, this no-go is intimately tied to these assumptions. Consequently, no long-lived trapped region can be described in this model, which suggests that additional structures are required to properly realize a model of Planck star as proposed in Ref. \cite{Rovelli:2014cta}. In particular, the lack of an inner horizon in the present model appears as the main missing ingredient.

Nevertheless, the present model provides a concrete framework to discuss the phenomenology of a UV complete pulsating star. Indeed, the interior geometry being modeled by a spatially closed bouncing universe, it enjoys both a UV and IR cut-offs denoted respectively $\tilde{\lambda}$ and $R_c$. As a result, the bouncing star follows cycles of expansion and contraction. This IR cut-off turns out to play a subtle role in this model. It encodes the minimal energy density $\rho_{\text{min}}$ of the star prior to collapse and contrary to the classical OS model where it is fixed by choosing the mass and maximal radius of the object, it is now a free parameter. Hence, compact objects are now described in a parameter space with one additional dimension, which allows one to consider at fixed mass and radius new much denser compact objects. This a major novelty of the present construction. Finally, the scale at which quantum gravity becomes non-negligible is encoded in the ratio $\tilde{\lambda}/ R_c$. While the UV cut-off $\tilde{\lambda}$ is expected to be universal and fixed once and for all, it is nevertheless possible to shift this scale thanks to the free IR cut-off, i.e by increasing the initial density $\rho_{\text{min}}$. This provides a concrete realization of the Planck star idea where the scale associated to quantum gravity effects depends crucially on the density of the compact object. Finally, we discuss the typical order of magnitude of the compact objects which can be described by our model, and show that they correspond to Planckian relics but not to standard macroscopic stellar objects. 

 To summarize, this work presents a minimal set up to implement the heuristic proposal of the Planck star using concrete techniques from LQC to account for the singularity resolution. Despite the no-go result discussed in Ref. \cite{paperUS}, which prevents the present model to discuss black-to-white hole bounce, the present construction shall serve as a platform to investigate the phenomenology of this pulsating star and as a guideline for further developments. More realistic assumptions might be introduced to evade the no-go and describe long-lived trapped region, such as a non-vacuum exterior geometry with an outer and inner horizons structure as presented in Ref. \cite{DeLorenzo:2014pta} for example. Moreover, more refined corrections to describe the interior dynamics could be used, such as inverse triad corrections \cite{Corichi:2013usa}.

This article is organized as follows. Section~\ref{section2} summarizes the effective construction developed in our companion article \cite{paperUS}. Then, Section~\ref{section3} describes the UV completion that arises from loop quantum gravity, revisiting and extending the results of the existing literature, among which analytic formula for the minimal radius at the bounce. Then, Section~\ref{section4} is devoted to the construction Planck star model. It presents the energy and pressure profiles of the thin shell, as well as its phase diagram, the dynamics of the star, its classical limit and the domain of applicability of our model.

\section{The model}\label{section2}

In this section, we present the exterior and interior geometries $\cM^{\pm}$ we shall glue across a time-like thin shell $\cT$. These two geometries describe the exterior and interior regions of the bouncing compact object. In each bulk region, we denote the four dimensional spacetime coordinates as $x_{\pm}^{\mu}$ with $\mu \in \{1,2,3,4 \}$ while on the thin shell, the three dimensional coordinates are denoted $y^a$ with $a \in \{ 1, 2, 3\}$. The space-like normal unit vector to the shell $\cT$ is denoted $n^{\pm}_{\mu} dx_{\pm}^{\mu}$ such that $n^{\mu}_{\pm} n^{\pm}_{\mu} = +1$. Finally, the projector from the bulk regions $\cM^{\pm}$ onto the shell $\cT$ is denoted $\; ^{\pm}e^{\mu}_{a} = \partial x_{\pm}^{\mu} / \partial y^a$. 


\subsection{Interior geometry}\label{sub2a}

Following the OS model, we shall assume that the interior geometry corresponds to a closed Friedmann-Lema\^{\i}tre (FL) universe. However, the goal being to model a UV complete gravitational collapse, we shall introduce modified Friedman equations in which the effective quantum corrections are parameterized in a general form. This will allow us to draw general conclusions independent of the specific form of the quantum corrections. 

Hence, the interior spacetime is assumed to have spatial sections that are homogeneous and isotropic with the topology of a $3$-sphere. Therefore, its metric is of the (FL) form
\begin{align}\label{FLmetric}
 \dd s^2_{-} & = \gamma^{-}_{\mu\nu} dx_{-}^{\mu} dx_{-}^{\nu}  \\
 & = - \dd\tau^2+  a^2(\tau) R^2_c \left[ \dd \chi^2 + \sin^2{\chi} \dd\Omega^2\right]
\end{align}
where $R_c$ is the constant curvature scale of the spatial sections, $\chi$ is the radial distance, in units of $R_c$, from the center and $a$ the scale factor describing the homologous evolution of the star. The dynamics of the interior region is assumed to follow some modified Friedmann equations to include the quantum corrections. In full generality, we assume that they take the form
\begin{align}
\label{HGen}
 \cH^2 & =\left( \frac{8\pi G}{3} \rho - \frac{1}{R_c^2a^2}\right)\left[ 1- \Psi_1(a)\right],\\
  \label{HdotGen}
 \dot{\cH} & = - 4 \pi G  (\rho+P)   \left[ 1 - \Psi_2(a)\right] +  \frac{1}{R_c^2 a^2} ,
\end{align}
where $\Psi_1$ and $\Psi_2$ are two functions that shall be specified by a choice of UV completion of GR and the Hubble function is defined as
\be
{\cal H}=\frac{\dot{a}}{a},
\ee
with a dot referring to a derivative with respect to $\tau$. Furthermore, we assume the conservation of the matter stress-energy tensor, i.e.
\begin{equation}\label{e.mcons}
 \dot{\rho} = -3{\cal H}(\rho+P).
\end{equation}
This choice of parametrization of the interior equations of motion, among which the classicality of the continuity equation, is well suited when modeling the interior dynamic with the LQC equations for the spatially closed universe. However, a more general set up might be useful for other models. 
In the case of dust ($P=0$), then
\be\label{CL}
 \rho =  \cE a^{-3} 
\ee
with ${\cal E}$ denoting the minimum of the energy density. Since at $\tau=0$, $a_0=1$ and $ \dot{a}_0$=0 , one obtains
\begin{equation}\label{e.7}
{\cal E}= \rho_{\text{min}} = \frac{3}{8\pi G R_c^2},
\end{equation}
which corresponds to the minimum value, $\rho_{\rm min}$, of the matter energy density. 
This allows one to rewrite the modified Friedmann equations in the more compact form
\begin{align}
 \cH^2R_c^2 &=\frac{1-a}{a^3}\left[ 1- \Psi_1(a)\right], \label{e.x1} \\
 R^2_c\dot{\cH} & = \frac{1}{a^2} -\frac{3}{2}\frac{1}{a^3}   \left[ 1 - \Psi_2(a)\right] ,\label{e.x2}
\end{align}
from which it follows, using Eq.~(\ref{e.mcons}), that
\be\label{e.xpsi2}
\Psi_2 = \left(1-\frac{2}{3}a\right) \Psi_1(a) - \frac{a(1-a)}{3}\frac{\dd \Psi_1}{\dd a}.
\ee
Finally, we note that the acceleration of the expansion is given by
\be
\frac{\ddot{a}}{a}R_c^2 = -\frac{1}{a^3}\left[\frac{1}{2} + \Psi_3(a) \right]  \;, \qquad \text{with} \qquad  \Psi_3(a) =(1-a)\Psi_1(a) +\frac{3}{2}\Psi_2(a).
\ee
Note that so far we have kept the form of the effective quantum corrections unspecified. Hence, the interior of the star is modeled by a closed FL universe with a modified dynamics encoded by the function $\Psi_1$. We can however draw some preliminary conclusions. If these corrections allow for a bounce, then there shall exist a time $\tau_b$ such that $\cH(\tau_b)=0$ and ${\cH}'(\tau_b)>0$. Generically, it implies that there shall exist a minimum value $a_{\rm min}\in ]0,1[$ such that $\Psi_1(a_{\rm min})=1$. Then, because of time reversibility and the absence of dissipation, this model will exhibit oscillatory solutions with $a\in ]a_{\rm min},1]$ and the resulting compact object will have a characteristic pulsation depending solely on the parameters of the effective quantum theory.

The time-like thin shell denoted $\cT$ in the following is chosen such that it corresponds to the surface of the star. We choose to fix its position at $\chi :=\chi_0$. This choice is not unique but appears quit useful to simplify the model. Then, the induced metric on the shell is given by
\be
\label{indmet1}
ds^2_{\cT} = \gamma^{-}_{ab}\dd y^a \dd y^b  = - \dd\tau^2+ R^2_c a^2(\tau) \sin^2{\chi_0} \,\dd\Omega^2
\ee
By construction, the unit normal vector is purely radial w.r.t. the interior coordinates,  $n^{-}_{\mu} \dd x_{-}^{\mu} = a(\tau) \dd\chi$. Observers co-moving with the shell have a $4$-velocity $u_{-}^\mu=(1,0,0,0)$ so that $u^{-}_{\mu} \dd x_{-}^{\mu} = \dd\tau$. Consequently, the proper time $\tau$ of the interior geometry coincides with the proper time of the rest frame of the shell $\cT$. The outer radius of the star evolves as
\be
\label{R}
R(\tau) = a(\tau) R_c \sin\chi_0
\ee
and $R_c$ has been chosen as unit of length. Since the star shall have a radius smaller than the spatial section, it can always be written as $\sin\chi_0$ if initially  we set $a_0=1$. Finally, the extrinsic curvature of the shell $\cT$ w.r.t the interior bulk region, i.e defined as $K^{-}_{ab} = e^{\mu}_a e^{\nu}_b\nabla_{\mu} n^{-}_{\nu}$ where $\nabla$ is the covariant derivative compatible with the 4d interior metric, is given by
\begin{align}
\label{ExCurv1}
K^{-}_{ab}\dd y^a \dd y^b & =   R_c a \sin\chi_0\cos\chi_0\, \dd\Omega^2.
\end{align}

\subsection{Exterior geometry}\label{sub2b}

For simplicity, we assume that the exterior of the star is modeled by the classical Schwarzschild geometry, such that the metric reads
\begin{align}
\dd s^2_+ & = \gamma^{+}_{\mu\nu} dx_{+}^{\mu} dx_{+}^{\nu}  \\ 
& =  - f(r) \dd t^2 + \frac{\dd r^2}{f(r)} + r^2 \dd \Omega^2  \;, \qquad \text{with} \qquad  f(r) = 1 - \frac{R_S}{r}
\end{align}
where $R_S$ is the Schwarzschild radius $R_S\equiv 2 GM$. As previously, we can always introduce $\chi_s$ such that
\be\label{e.defchiS}
R_S= 2 GM = R_c\sin\chi_s.
\ee
This choice for the exterior geometry provides the minimal set up to mimic the classical Oppenheimer-Snyder construction and is motivated by simplicity. While one should then generalize this construction to include deviations form the classical Schwarzschild geometry as a next step, we focus here on this minimal set up as a preliminary step. See Ref.~\cite{next} for such an extension. Moreover, this choice of exterior geometry will allow us to draw interesting conclusions regarding the allowed energy scale of the bounce. Indeed, the presence of a single horizon in the exterior geometry will have some important consequences on the bouncing compact object. 
 
Let us now discuss the embedding of the time-like thin shell $\cT$ in this exterior geometry. On the shell $\cT$ describing the outer boundary of the star, the time and radial exterior coordinates can be written as $t := T_{+} (\tau)$ and $r := R_{+}(\tau)$. Then the induced metric on $\cT$ is given by
\be
\label{indmet2}
ds^2_{\cT} = \gamma^{+}_{ab}\dd y^a \dd y^b = - \left[ f \dot{T}^2_{+} - \frac{\dot{R}^2_{+}}{f} \right] \dd \tau^2 + R^2_{+}(\tau) \dd\Omega^2
\ee
The normal spacelike vector $n^{\mu} \partial_{\mu}$ to $\cT$ is given by
\be
n^{\mu}\partial_{\mu} = \frac{\dot{R}}{f} dT + f \dot{T} dR \;, \qquad n_{\mu} dx^{\mu} = - \dot{R} dT + \dot{T} dR
\ee
such that $n^{\mu} u_{\mu} =0$ where  $u^{\mu}\partial_{\mu} = e^{\mu}_a u^a \partial_{\mu} $ with $u^a \partial_a = \partial_{\tau}$ the three dimensional tangent time-like unit vector  associated to an observer at rest on $\cT$.
 The extrinsic curvature of $\cT$ with respect to the exterior bulk region $\cM^{+}$, in term of the coordinates $(\tau, \theta, \phi)$ on $\cT$, is given by
\be
\label{ExCurv2}
K^{+}_{ab} dy^a dy^b = - \frac{1}{ f \dot{T}} \left[ \ddot{R}   + \frac{f_{,R}}{2} \right]  d\tau^2 +  \dot{T} f R d\Omega^2
\ee 
This concludes the presentation of the exterior and interior geometries we will use to describe our bouncing compact object.

\subsection{Junction conditions}\label{sub2c}
\label{main}
We need now to specify the junction conditions to be imposed on the shell $\cT$. In principle, these matching conditions depend on the specific theory under consideration. However, when considering bouncing geometries descending from LQC, as we shall do in the last section, there is no known covariant lagrangian one could start with to derive the associated junction conditions. Therefore, in order to make progress, one can use the freedom to interprete the interior geometry as a GR solution with some exotic energy-momentum tensor. Then, one can use the standard Israel-Darmois junction conditions to perform the gluing \cite{D, Israel:1966rt,Barrabes:1991ng}. This is the strategy we shall follow in this work.

Consider therefore a perfect fluid on the time-like shell $\cT$ whose energy-momentum tensor $\cS_{ab}$ takes the form
\be
\cS_{ab} = \sigma u_a u_b + p \left( \gamma_{ab} + u_a u_b \right)
\ee
where $u_a = \left( 1,0,0\right)$ is the $3$-velocity of the observer at rest with the fluid. Then, one has that $\cS^a{}_b = \text{diag}\left( -\sigma, p, p\right)$. The ID junction conditions to imposed are given by
\begin{align}
\label{Kin}
[\gamma_{ab}] & = 0 \\
\label{Dyn}
[K_{ab} - \gamma_{ab} K] &= - 8\pi G \cS_{ab}
\end{align}
where $\cS_{ab}$ is the stress energy-momentum tensor on the time-like shell $\cT$ and where the bracket of a quantity $X$ corresponds to the discontinuity of this quantity across $\cT$, i.e to $[X] = X_{+} - X_{-} |_{\cT}$. The first one imposes the continuity of the induced metric across the shell $\cT$, and is thus purely kinematical. The second conditions provides the equation of motion of the shell. Moreover, these two conditions together with the GR scalar and momentum constraints imposed on the shell $\cT$ automatically imply two additional conservation equations. However, being a consequence of the other equations, these two equations do not impose additional constraints and we do not write them here. See \cite{Israel:1966rt,Barrabes:1991ng} for details.

Using (\ref{indmet1}) and (\ref{indmet2}), the kinematical constraints (\ref{Kin}) read
\begin{eqnarray}
\label{condkin}
 f \dot{T}^2_{+} - \frac{\dot{R}^2_{+}}{f} = 1 \;, \qquad R_{+}(\tau) = R_c a(\tau) \sin{\chi_0}
\end{eqnarray}
For now on, we shall write $R_{+}(\tau) = R(\tau)$ for simplicity. Finally, in order to solve for the dynamical condition (\ref{Dyn}), we introduce the notation $\kappa^a{}_b = [K^a{}_b - \delta^a_b K]$ which allws one to write
\begin{align}
\sigma = - \frac{\kappa^{\theta}{}_{\theta}}{4\pi G} \;, \qquad p = \frac{1}{8\pi G} \left( \kappa^{\tau}{}_{\tau} + \kappa^{\theta}{}_{\theta}\right)
\end{align}
Using (\ref{ExCurv1}) and (\ref{ExCurv2}), one obtains that
\begin{align}
\label{en}
\sigma & = \frac{1}{4\pi G R} \left( \cos{\chi_0} - f \dot{T} \right) \\
\label{press}
\sigma + 2p & =  \frac{1}{4\pi G  f \dot{T}} \left[ \ddot{R}   + \frac{f_{,R}}{2} \right] 
\end{align}
This fixes the profiles of the surface and pressure of the thin shell. 

\subsubsection{Extended mass relation}\label{sub2d}

The exterior and interior geometries involve two constants of motion $M$ and ${\cal E}$. Starting from (\ref{en}) and upon using the modified Friedman equations (\ref{e.x1}) as well as the kinematical constraints (\ref{condkin}), one obtains a dynamical relation relating these two constants of motion and the surface energy of the thin shell. Introducing the quantity $\Sigma = 4\pi G R_c \sigma$, this key relation is given by
\begin{align}
\label{Master-Equation}
M = \frac{4\pi }{3} \rho R^3 -  \frac{R^3}{2G R^2_c} \left[ \Sigma^2 - \frac{2 R_c \cos{\chi_0 }}{R} \Sigma +  \frac{ 1-a}{a^3} \Psi_1 \right],
\end{align}
which can be recast, using (\ref{e.defchiS}) and (\ref{R}), into 
\begin{align}
\label{e.massrel}
\sin\chi_s &=  \sin^3\chi_0\, \left[ 1 - a^3\left( \Sigma^2 - \frac{2\cot\chi_0}{a}\Sigma + \frac{ 1-a}{a^3}\Psi_1
\right)
\right],
\end{align}
which depends explicitly on the effective quantum correction $\Psi_1$. Let us discuss this matching constraint.
First of all, in GR, i.e. when $\Psi_1=0$, the junction condition is possible without mass shell only if the standard OS condition is satisfied, 
\be
\label{equality}
\sin\chi_s=\sin^3\chi_0
\ee 
which just means that $M=4\pi\rho(\tau) R^3(\tau)/3$. Hence, the standard mass relation obtained in the  OS model is recovered if $\Psi_1=0$ and $\Sigma=0$. It follows that otherwise (i.e. departure from GR or mismatch between the collapsing mass and exterior mass) the exterior and interior geometries can be glued only at the price of a shell. This extended mass relation turns out to play a key role in that it ensures that the conserved quantities associated to the exterior and interior geometries properly match during the whole evolution of the compact object, even during the bounce. 

An interesting outcome of this extended mass relation is that the energy of the dust $\cE$, which is related to the IR cut-off, is now a free parameter of the model, contrary to the classical OS model where it is fixed by the mass $M$ and the maximal radius $R_{\text{max}} = R_c \sin{\chi_0}$. Hence, once the UV cut-off $\lambda$ is fixed once and for all\footnote{The Barbero-Immirzi parameter does not play any relevant role and can also be fixed to $\gamma=1$ for simplicity}, the star is not parametrized by two but three independent parameters $\left( R_c, \chi_s, \chi_0\right)$. At fixed $\left( \chi_s, \chi_0\right)$, one has now the freedom to shift the density of the compact objects described by the model. This new freedom is a direct consequence of the presence of the thin shell which allows one to split the total mass between the bulk interior region and the thin shell. It is then interesting to investigate how each physical quantity and scale entering in the model behave under a rescaling of the minimal energy density $\rho_{\text{min}}$, or equivalently $R_c$. From the very definition of the Schwarzschild radius, it is clear that a rescaling $R_c \rightarrow \hbox{e}^{\epsilon} R_c$ implies a rescaling of the mass $M \rightarrow \hbox{e}^{\epsilon} M$. We show in Appendix~\ref{scaling} that the generalized mass relation is invariant under such rescaling, as expected. An interesting consequence of this mass rescaling is that the mass being canonically conjugated to the asymptotic Killing time in spherical symmetry \cite{Kuchar:1994zk}, one has to further rescale this Killing time in order to keep the symplectic potential invariant. This point should be investigated further but it goes beyond the scope of this work and we should not comment further on this.

\subsubsection{Constraint  of the thin shell dynamics}\label{sub2e}

From the previous junction conditions have been obtained, we can now find suitable physical solution for the profile of the energy and pressure of the thin-shell. The relation~(\ref{e.massrel}) gives a second order polynomial equation for $\Sigma$,
\begin{eqnarray}
\label{sigm}
\Sigma^2 - \frac{2\cot\chi_0}{a}\Sigma + \frac{1}{a^3}
\left[ \frac{\sin\chi_s}{\sin^3\chi_0} -1 + (1-a)\Psi_1
\right]=0. \nonumber
\end{eqnarray}
This sets a condition on the parameters of the star $(\chi_0,\chi_s)$ and of the effective quantum theory through the dependency on $\Psi_1$. From this polynomial second order equation, one obtains a dynamical condition for the collapse, given by 
\begin{eqnarray}\label{e.det00}
\Delta &=&\frac{4}{a^2}\left\lbrace\cot^2\chi_0  - \frac{1}{a}\left[ \frac{\sin\chi_s}{\sin^3\chi_0} -1 + (1-a)\Psi_1
\right] 
\right\rbrace \geqslant 0
\end{eqnarray}
which implies that the discriminant of (\ref{sigm}) is positive for all allowed values of $a$. Provided this is the case, the surface energy of the thin shell admits two branches of solutions given by
\begin{eqnarray}\label{e.defSig1}
\Sigma_\pm &=& \frac{\cot\chi_0}{a}\pm \frac{1}{2}\sqrt{\Delta} .
\end{eqnarray}
Since the thin shell shall be understood as partially induced by quantum effects, we shall allow it to violate the standard energy conditions, and in particular allows the energy profile to be negative. For that reason, we will focus on the minus branch $\Sigma_{-}$ when building the model in the last section. Let us finally point that this solution to the junction condition has been obtained without specifying the quantum correction $\Psi_1$.

Let us finally point that the profiles of the energy and pressure of the thin shell depend explicitly on the quantum correction $\Psi_1$ and its derivatives. Hence, the thin-shell can be understood as an effective way to encode part of the quantum effects for such a gluing. As a result, the quantum corrections will affect the thin-shell. If the surface of the collapsing compact object crosses its Schwarzschild radius to form an event horizon, and if the quantum effects remain negligible at the horizon formation threshold and are triggered only near the would-be singularity, then, they could be considered as confined to the deep interior. However, as we are going to see now, consistency of the matching leads to a radically different picture, where the bounce turns out to occur prior to horizon formation, or at most, precisely at the horizon formation threshold.

\subsection{Bound on the energy scale of the bounce}

One major question for such bouncing compact object is at which energy scale the quantum effects become dominant? Or in other word, at which energy scale the bounce occurs ? It is widely believed that quantum gravity effects shall only be relevant in the deep interior region near the would be singularity. 

As it turns out, a surprising outcome of our construction is to provide a constraint on the energy scale at which a bounce can occur. Let us summarize the main argument of Ref.~\cite{paperUS}. First, let us introduce the lapse function $A[\tau] = 1/\dot{T}$. Using the kinematical condition
\be
 f \dot{T}^2 - \frac{\dot{R}^2}{f} \big{|}_{\cT}= 1
\ee
and that the time derivative of the dynamical radius reads $\dot{R} = R(\tau) \cH(\tau)$ where $\cH$ is the Hubble factor associated to the interior region, the expression of the lapse $A[\tau]$ can be recast in the more suggestive form
\be
\label{keyrel}
A^2[\tau] =  \frac{1}{\dot{T}^2} = \frac{f^2[R(\tau)]}{f[R(\tau)] + R^2(\tau) \cH^2(\tau)} \;, \qquad \text{with} \qquad f[R(\tau)] = 1 - \frac{R_s}{R(\tau)}.
\ee
It is then straightforward to see that if a bounce occurs, i.e. if at some time $\tau_b$, one has $\cH(\tau_b)=0$ and $\cH'(\tau_b)>0$, then the lapse at the bounce satisfies
\be
\label{KEY}
A^2(\tau_b)= 1 - \frac{R_s}{R(\tau_b)} \geqslant 0.
\ee
For a star initially above the Schwarzschild radius, consistency imposes that 
\be
\label{kkin}
R(\tau_b) \geqslant R_s
\ee 
during the collapse and the bounce. This implies that the star never crosses its Schwarzschild radius such that it prevents a trapped region to form. To finish, note that while we are indeed working with a singular coordinate system at the horizon, the same computation can be done in the Eddington-Finkelstein coordinates, regular at the horizon, which leads to the very same result \cite{paperUS}. Notice also that this result is not contradictory with the bouncing black hole model built in Ref.~\cite{Haggard:2014rza}, in which a collapsing null shell was considered. In that case, the key relation (\ref{keyrel}) responsible of our result is not present.

Finally, let us make one more remark. Assuming a bounce from quantum origin for the interior geometry, the above result might misleadingly suggest that quantum gravity effects are triggered at low curvature, which would go against the intuitive expectation. It turns out that this is not the case.
In Section~\ref{om}, we shall show that when the interior dynamics is modeled by a quantum bounce descending from loop quantum cosmology, consistency conditions severely restrict the range of applicability of the model.
As it turns out, it can be shown that macroscopic stellar objects are excluded by this model, and only Planckian relics can be consistently considered. For such extremely small mass and size objects, curvature is already very high even outside the horizon. As a consequence, even if the bounce occurs above or at the horizon scale, it still occurs in a regime of high curvature, as intuitively expected from a quantum bounce.

\subsection{Summary}

Without specifying the effective quantum corrections which ensure the UV completion of the interior dynamics, we have shown that one can still derive a general solution of the Israel-Darmois junction conditions, providing an effective theory for a UV complete gravitational collapse. The consistency of the model is encoded in the generalized mass relation (\ref{e.massrel}). As a result, one obtains a general profile for the surface energy $\Sigma$ and surface pressure $\Pi$ of the thin shell. Before discussing a concrete realization for the UV complete interior, let us recall the different assumptions of this construction. Following the thin shell approach developed in Ref.~\cite{paperUS}, 
\begin{itemize}
\item the collapsing star is modeled as a closed FL universe with a modified UV complete dynamics. The UV completeness is encoded in the correction $\Psi_1$ which shall descend from the UV completion of GR. This implies that the dynamics of the scale factor can describe either a {\em bouncing star} or a {\em bouncing universe} and that it will remain independent, for the former, of the properties of the thin shell.
\item The exterior of the star is the standard vacuum Schwarzschild geometry. 
\item The two geometries are glued on a time-like thin shell. Independently of the solution for the expansion $a(\tau)$,  the properties of the thin shell are fully determined by the Israel-Darmois junction conditions. They depend on both the characteritic of the star (initial mass $\chi_s$ and initial radius $\chi_0$) and on the function $\Psi_1$ which contains the parameters of the quantum theory. It follows that the physically acceptable solutions are constrained to satisfy the mass relation~(\ref{e.massrel}). 
\end{itemize}
We have concluded that if $\Psi_1$ allows for a singularity resolution which manifests through a bounce of the star, then the compact object experiences cycles of contraction and expansion without black hole formation. This provides the basics of an effective theory of a pulsating Planck star. To go further one needs to choose $\Psi_1$.

\section{UV completion from Loop Quantum Cosmology}\label{section3}

To be more specific, we now assume that $\Psi_1$ is determined by the corrections that arise in the loop regularization of the dynamics of spatially spherical universe presented in Refs.~\cite{Ashtekar:2006es, Szulc:2006ep, Corichi:2011pg, Dupuy:2016upu, Corichi:2013usa}. Two different loop quantization schemes have been discussed in the literature, the so called curvature regularization~\cite{Ashtekar:2006es, Szulc:2006ep}, and the connection regularization~\cite{Corichi:2011pg, Dupuy:2016upu}, while additional refinement related to inverse volume corrections have been discussed in  Ref.~\cite{Corichi:2013usa}. More recently, new phenomenology of this bouncing cosmology has been discussed in Ref. \cite{Dupuy:2019ibu}. See also Ref. \cite{Singh:2013ava} for another regularization scheme. In what follows, we focus our attention on the connection regularization scheme, the curvature regularization scheme phenomenology will be discussed elsewhere.

\subsection{Connection regularization - overview}

Let us start by the modified Friedmann equations obtained in Ref.~\cite{Corichi:2011pg} and discuss their consequences for our model. While this model has already been investigated in Refs.~\cite{Corichi:2011pg, Dupuy:2016upu}, the present section provides new analytic results regarding the expression of the minimal radius, as well as a clearer picture of the oscillating dynamics, and in particular the existence of two cycles of bounces associated to the two different minimal radii. Consider the metric of a closed FL universe
\be
 \dd s^2 = - N^2 \dd \tau^2 + R^2_ca^2(\tau) \left(\dd \chi^2 + \sin^2{\chi} \dd \Omega^2\right).
\ee
Let us briefly summarize how the modified Friedmann equations are obtained within the LQC framework. Following the notations of Ref. \cite{Ashtekar:2006es},
the gravitational canonical variables are given by
\be\label{e.defvb}
b \equiv \frac{c}{\sqrt{|p|}},\qquad v \equiv |p|^{3/2}, \;\;\; \text{such that} \;\;\; \{ b, v\} = 4\pi G \gamma,
\ee 
which correspond respectively to the Hubble factor and the 3-volume of the spatially closed universe. The dimension are given by $[b] = L^{-1}$, $[v] = L^3$. 
Then, using  the loop regularization of the phase space of the closed FL universe~\cite{Corichi:2011pg}, the modified dynamics is encoded in the polymer Hamiltonian constraint
\be\label{hfulll}
\cS_{\text{full}}[N]  = N \left( v \rho  - \frac{3}{8\pi G \gamma^2 \tilde{\lambda}^2} v  \left[ (\sin \tilde{\lambda} b - D)^2+ \gamma^2 D^2 \right]\right) \simeq 0
\ee
where $\rho$ is the energy density, and we have  introduced the short notation $D = \kappa /v^{1/3}$, hence related to the scale factor by
\be
D(\tau) = \frac{\tilde{\lambda}}{R_c a(\tau)}.
\ee
It follows that the effective quantum theory depends on two parameters. The UV cut-off is encoded in the parameter $\tilde{\lambda}$, with dimension of length, i.e $[\tilde{\lambda}] =L$, while $\gamma$ is the Barbero-Immirzi parameter. without dimension. It is worth keeping in mind that these parameters are not fixed a priori, and shall be determined by experiments. 
The equations of motion w.r.t. the cosmic time $\tau$, i.e. $N=1$, are then given by
\begin{align}
\dot{v}  &  = \frac{3}{\gamma \tilde{\lambda}} v \left( \sin\tilde{\lambda} b - D \right)\cos\tilde{\lambda} b \label{e.dyn1}\\
\dot{b} & = - \frac{3}{2\gamma \tilde{\lambda}^2} \left[ \sin^2 \tilde{\lambda} b - \frac{4D}{3} \sin \tilde{\lambda} b  + \frac{1}{3} (1+\gamma^2) D^2 \right]\label{e.dyn2}
\end{align}
where we have assumed that $P = \partial_v \left( v \rho \right)=0$, i.e. we work with a dust field of matter. The scalar constraint~(\ref{hfulll}) can be recast as
\begin{equation}\label{e.cc0}
 \frac{\rho}{\rho_c}=  (\sin \tilde{\lambda} b - D)^2+ \gamma^2 D^2, 
\end{equation}
where we have introduced the critical energy density
\begin{equation}
\rho_c \equiv \frac{3}{8\pi G \gamma^2 \tilde{\lambda}^2} = \frac{{\cal E}}{\gamma^2(\tilde{\lambda}/R_c)^2}.
\end{equation}
This concludes our description of the effective phase space and the associated effective dynamics of the closed FL geometry filled up with a perfect pressureless fluid. 

\subsection{Evolution of the quantum correction $\Psi_1$}

From now on, we redefine the variables such that $\tau \rightarrow \tau/R_c$, and $b\rightarrow  R_c b$. Rescaling also the UV cut-off the same way, we introduce the new dimensionless parameter
\be
\lambda = \frac{\tilde{\lambda}}{R_c}
\ee
which shall play a crucial role in the following. Being the ratio of the UV and IR cut-offs, it encodes the scale at which quantum gravity effects become non-negligible. The interesting point is that the IR cut-off $R_c$ is actually related to the energy density of the compact object prior to collapse, through the relation (\ref{e.7}). Small value of $R_c$ corresponds to high energy density. Now, while it is expected that the UV cut-off $\lambda$ is a universal quantity fixed once and for all for any star, the parameter $R_c$ is now a free parameter of the model which characterizes the star just as its mass and radius. Hence, by shifting the density of the compact object at fixed mass and radius, i.e at fixed $\left( \chi_s, \chi_0\right)$, one can shift the effective scale of quantum gravity effects.  This allows to implement the idea advocated in Ref. \cite{Rovelli:2014cta}, that provided a compact object reaches a sufficiently large density w.r.t the Planck density, quantum gravity can be triggered even if the size of the object is much larger than the Planck length. In the case of the present model, this interpretation is possible thanks to the double role played by the scale $R_c$, encoding both the spatial curvature of the interior geometry and the minimal energy density of the compact object.  

Now, the above rescaling also implies that $D=\lambda/a$. Finally, it is convenient to introduce the rescaled Hubble factor and the rescaled critical energy
\begin{equation}\label{e.311}
 H = {\cal H}R_c \;, \qquad \rho_c \equiv \frac{{\cal E}}{\gamma^2 \lambda^2}.
\end{equation}
Since $v\propto a^3$, Eq.~(\ref{e.dyn1}) implies that
\begin{equation}
H=\frac{1}{\gamma\lambda} \left( \sin\lambda b - D \right)\cos\lambda b.
\end{equation}
This gives an expression for $H^2$ in which one can express $(\sin\lambda b - D)^2$ thanks to Eq.~(\ref{e.cc0}) as
\begin{equation}\label{e.temp}
\left( \sin\lambda b - D \right)^2=\gamma^2\lambda^2\left(\frac{8\pi G}{3}\rho-\frac{1}{a^2}\right).
\end{equation}
It follows from the definition~(\ref{HGen}) that $1-\Psi_1=\cos^2\lambda b$, from which we conclude that
\begin{eqnarray}
\label{qcor}
\Psi_1 & = & \sin^2\lambda b. \label{cor1}
\end{eqnarray}
This expression shows that $\Psi_1=1$ when $\lambda b =\pm\pi/2$. It follows that the first condition for the existence of a bounce is satisfied.

\begin{figure}[h]
 \centering
   \includegraphics[scale= .45]{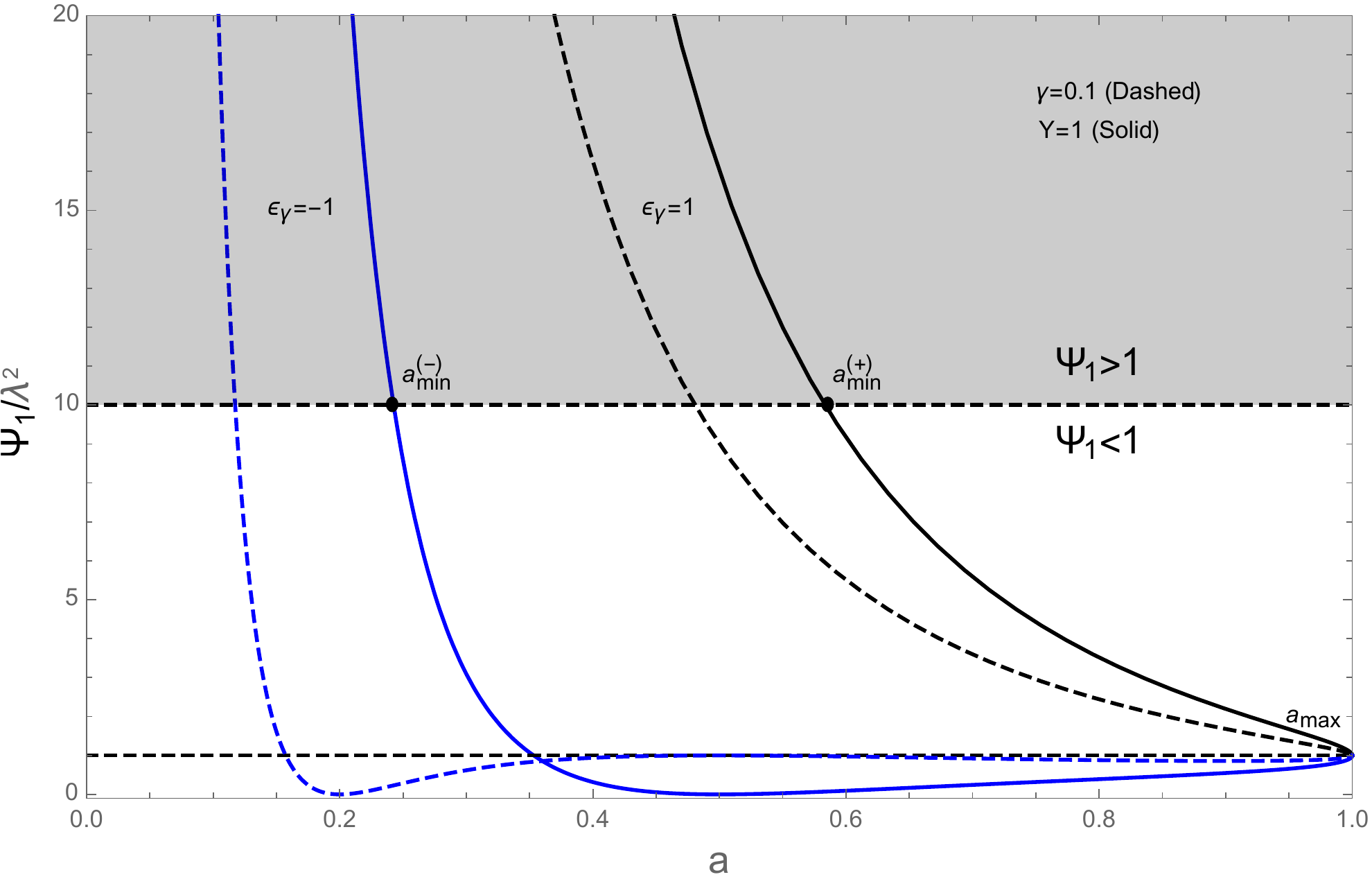}
   \caption{$\Psi_1(a)$ in units of $\lambda^2$ (black: $\varepsilon=+1$, blue: $\varepsilon=-1$) for $\gamma=1$ (solid) and $0.1$ (dashed). If $\lambda<1$, there exists two values, $a_{\rm min}^{(-)}<a_{\rm min}^{(+)}$ for which $\Psi_1=1$, i.e. $H=0$.}
   \label{fig:psi1} 
\end{figure}

It can also be expressed in terms of $a$ by using the expression for $\rho$ and making use of Eq.~(\ref{e.7}) in Eq.~(\ref{e.temp}) to express $\sin\lambda b$, leading to
\be\label{e.Psi1deA}
\Psi_1 = \frac{\lambda^2}{a^2}\left( 1 + \varepsilon \gamma \sqrt{\frac{1}{a} - 1} \right)^2
\ee
where $\varepsilon=\pm1$ is a sign that we shall discuss later. This expression allows us to get some insight on the dynamics.  Fig.~\ref{fig:psi1} shows the existence of two branches labeled by the sign of $\varepsilon$. If $\lambda<1$, there will always exist two values,
$$
0<a_{\rm min}^{(-)}<a_{\rm min}^{(+)}<1,
$$
such that $\Psi_1(a)=1$. By definition, $\Psi_1[a_{\rm min}^{(\pm)}]=1$ corresponds to  a vanishing Hubble factor, $H=0$. Hence, we expect to have two different kinds of bounce. This was first pointed out in Ref. \cite{Corichi:2011pg}.

Following Ref.\cite{Corichi:2011pg}, it is useful to write down the modified Friedmann equations in the more familiar form
\begin{align}
\label{HLQCC11}
\cH^2 & = \frac{8\pi G}{3} \left( \rho - \frac{1}{R_c^2a^2} \right) \left( 1 - \frac{\rho- \rho_1}{\rho_c}\right) \\
\label{HLQCC22}
\dot{\cH} + \frac{1}{R_c^2a^2} & = - 4 \pi G \left( \rho - \frac{2\rho_1}{3} \right) \left( 1 - \frac{\rho- \rho_2}{\rho_c}\right) 
\end{align}
in terms of the densities
\begin{align}
\label{e.defrhoo}
  \rho_1  &= \rho_c D \left[ (1+\gamma^2) D - 2 \sin{(\lambda b)} \right]  \\ 
  \rho_2 & = \rho_c D \left[ (1+\gamma^2) D -  \sin{(\lambda b)} \right].
\end{align}
these expressions allow us to extract $\Psi_1$ and $\Psi_2$ in term of the density $\rho$ and the critical densities $\rho_1$ and $\rho_2$ and $\rho_c$. Comparing with Eq.~(\ref{HGen}), one obtains
\begin{align}
\Psi_1 = \frac{\rho- \rho_1}{\rho_c}
\end{align}
which can be checked to be equivalent to the expressions~(\ref{cor1}) and~(\ref{e.Psi1deA}). To finish, $\Psi_2$ is obtained from Eq.~(\ref{e.xpsi2}). As a final remark, it is straightforward to check that the classical limit corresponds to 
\be
\lambda = \frac{\tilde{\lambda}}{R_c} \rightarrow 0\;, \qquad \Rightarrow \qquad \rho_c \rightarrow +\infty\;, \qquad D \rightarrow 0 \;, \qquad \Psi_1 \rightarrow 0
\ee 
where we have used Eq.~(\ref{e.defrhoo}). 
\subsection{Dynamics of the oscillating closed universe (interior region)}

\label{interiornum}

As already discussed in Ref. \cite{Corichi:2011pg}, the bouncing dynamics encoded in the modified Friedmann equations (\ref{HLQCC11}-\ref{HLQCC22}) and (\ref{CL}) enjoys two bounces with two different allowed minimal radii. In order to understand the dynamics, it is useful to recast the equations of motion in a more suitable form. In the following, we provide the expression of the two allowed minimal radii and discuss the different cycles experienced by this bouncing closed universe.

\subsubsection{Reduced form of the dynamical equations}
\label{RFDYN}
It turns out to be convenient to introduce the variables
\be\label{e.cvar}
 u\equiv1/a, \qquad X\equiv\sin\lambda b.
\ee
Concerning  Eq.~(\ref{e.dyn1}), one uses Eq.~(\ref{e.temp}) with the expression~(\ref{CL}) for the density to eliminate the quadratic term $\sin^2\lambda b$ to finally get
\begin{eqnarray}
\frac{\dd u}{\dd\tau} &=&-\frac{u}{\lambda\gamma}\cos\lambda b \left(X-\lambda u\right)  \label{e.s1}\\
\frac{\dd X}{\dd\tau} &=& -\lambda\gamma u \cos\lambda b \left\lbrace 
\frac{3}{2}u^2-\frac{\left[(1+\gamma^2)\lambda u -X
\right]}{\lambda\gamma^2}
\right\rbrace\label{e.s2}
\end{eqnarray}
with 
\be
\cos\lambda b = \pm \sqrt{1-X^2}.
\ee
This latter sign has no influence on the dynamics since it can be absorbed by the redefinition $\tau\rightarrow-\tau$.
Additionally to this set of two equations of motion, the on shell system satisfies the constraint~(\ref{e.temp}) that takes the form
\be\label{e.c00}
\left( X - \lambda u \right)^2 = \gamma^2\lambda^2 u^2 \left(u-1\right),
\ee
which corresponds to satisfy the scalar constraint during the whole evolution. Despite the simplicity of this set of equations, one cannot obtain a closed analytical solution for $u(\tau)$ and one has to numerically integrate this set of coupled equations of motion. This point is explained in detail in Appendix~\ref{B}.

\subsubsection{Minimal radius}

\begin{figure}[h]
 \centering
   \includegraphics[scale= .4]{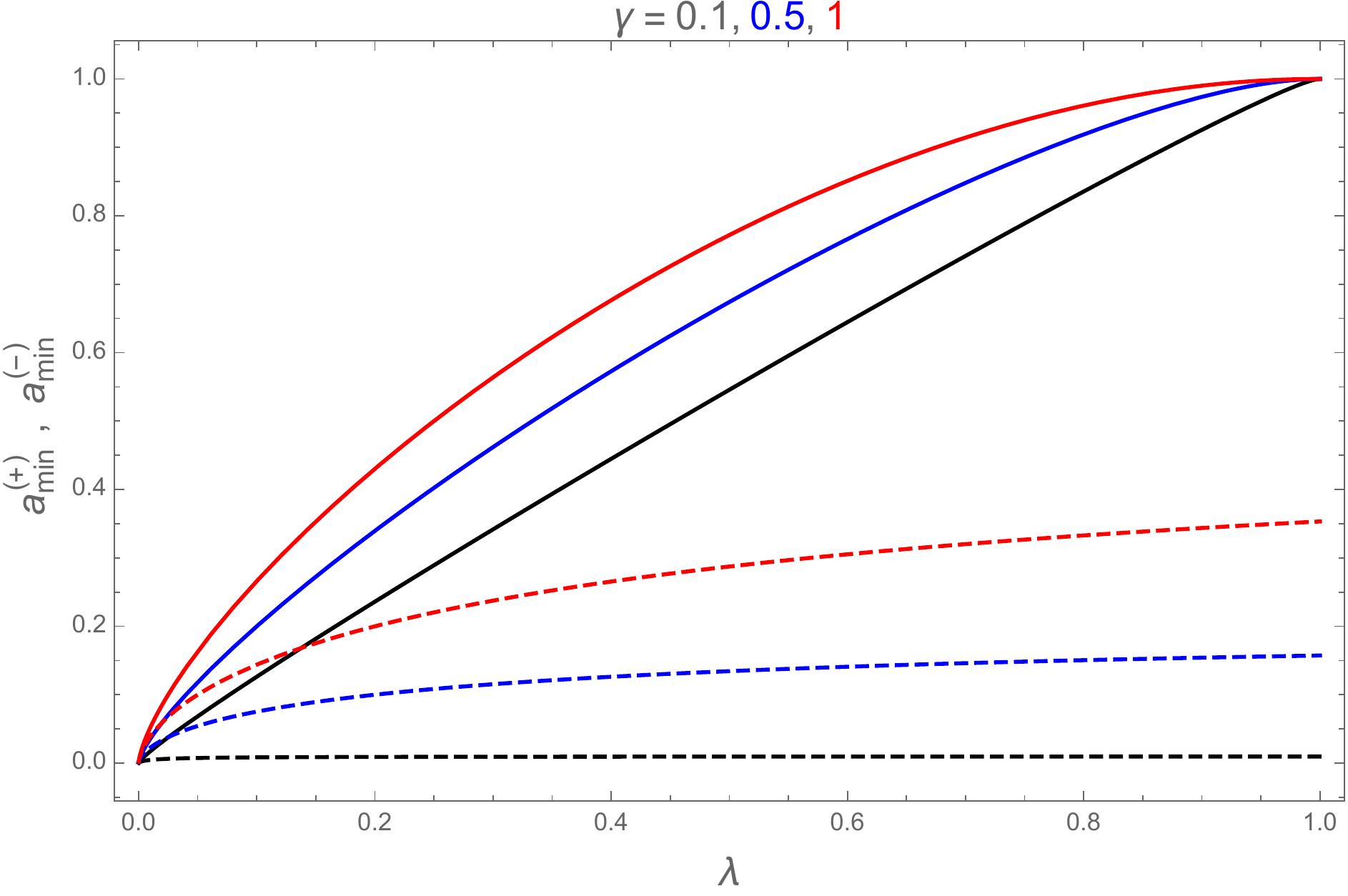}
   \caption{The evolution of $a_{\rm min}^{(+)}$ (solid) and $a_{\rm min}^{(+)}$ (dashed) with $\lambda$. }
   \label{fig:amin} 
\end{figure}

As seen from Fig.~\ref{fig:psi1}, we expect the dynamics to enjoy two minimal radii. Indeed the system~(\ref{e.s1}-\ref{e.s2}) has 3 configurations in which $\dd a/\dd\tau =0$. The first one is obviously characterized by $u=1$ (i.e. $a=1$), which corresponds to the maximal extension of the star. It corresponds to the initial state which defines the initial conditions to integrate the system~(\ref{e.s1}-\ref{e.s2})
\be\label{e.condiniti}
 u=1, \qquad a=1, \qquad X= \lambda.
\ee
The second set of solutions is obtained for $\cos\lambda b=0$, that is for
\be\label{e.cyclen}
\lambda b = \frac{\pi}{2}+ n\pi,\qquad n\in\mathbb{Z},
\ee
so that $\sin\lambda b=(-1)^n$. Inserting this condition in the effective Friedmann equation leads to
\be\label{e.tp1}
\lambda u \left[1+\varepsilon \gamma\sqrt{u-1}\right]=(-1)^n.
\ee 
Since $u>1$ (i.e. $a<1$), we can define $y$ such that $u=1+y^2$,  so that the solutions of  Eq.~(\ref{e.tp1}) are the roots $y_*(n,\lambda,\gamma)$ of
$$
\lambda (y^2+1)(1+\varepsilon \gamma y)=(-1)^n.
$$ 
First, note that changing the sign of $\varepsilon$ changes the sign of the root of this equation, which has no influence on the value of the physical variables $u$ or $a$. So we can arbitrarily choose $\varepsilon=-1$. Then the solution with $n$ odd and even are related by the change of the sign of $\lambda$. It follows that the solutions are $y_*(\lambda,\gamma)$ for $n=2p$ and $y_*(-\lambda,\gamma)$ for $n=2p+1$ with $y_*$ the root of
\be
\label{y}
 \lambda (y^2+1)(1- \gamma y)=1
\ee
with $y>0$. The discriminant of this third order polynomial equation is given by
\be
\Delta_O = 4 \lambda^2 \left[ \left( 9\gamma^2 + 1\right) \lambda - \frac{27}{4} \gamma^2 - \left( 1 + \gamma^2 (2+\gamma^2)\right) \lambda^2 \right].
\ee
The nature of the roots (either complex or real) depends on the sign of $\Delta_O$. In our case, the parameter $\gamma$ is of order unity while $ \lambda \ll 1$ for a realistic astrophysical object. Therefore, the first and last terms are very small compared to the middle term and one has therefore
\be
\Delta_O\simeq - 27\gamma^2\lambda^2 <0.
\ee 
This implies that the third order polynomial equation (\ref{y}) has only one real root $y_{\ast}$ supplemented with two complex conjugated roots that are therefore not physical. The only real root is
\be
 y_* = \frac{1}{3\gamma}\left[1+\frac{2^{1/3}(3\gamma^2-1)}{\delta_2} - \frac{\delta_2}{2^{1/3}}\right]
\ee
with
\begin{eqnarray}
 \delta_2&=&\left(-2-18\gamma^2+27\frac{\gamma^2}{\lambda}+\delta_1 \right)^{1/3}, \\
 \delta_1&=&\sqrt{4(3\gamma^2-1)^3-\left(2+18\gamma^2-27\frac{\gamma^2}{\lambda}\right)}.
\end{eqnarray}
In conclusion, irrespective of the sign $\varepsilon$, the scale factor $a$ enjoys two minima
\be\label{e.defamin}
 a_{\rm min}^{(\pm)}(\gamma,\lambda)\equiv\frac{1}{y_*^2(\gamma,\pm\lambda)+1},
 \ee
the sign of $\lambda$ being related to the phase of $\lambda b$ that determines the sign of $X$. These minima satisfy $0<a_{\min}^{(-)}<a_{\min}^{(+)}<1$ and they scale as
\be\label{e.scaling}
a_{\rm min}^{(\pm)}\simeq \gamma^{2/3}\lambda^{2/3} \pm \frac{2}{3}\lambda\quad \hbox{as} \quad \lambda\rightarrow 0.
\ee
Fig.~\ref{fig:amin} depicts the dependence of these minima with $\lambda$ for different values of the parameter $\gamma$. As expected from Eq.~(\ref{e.Psi1deA}), $a_{\rm min}^{(+)}\rightarrow 1$ when $\lambda\rightarrow 1$.

\subsubsection{Cycles of bounces}

\begin{figure}[h]
 \centering
   \includegraphics[scale= .5]{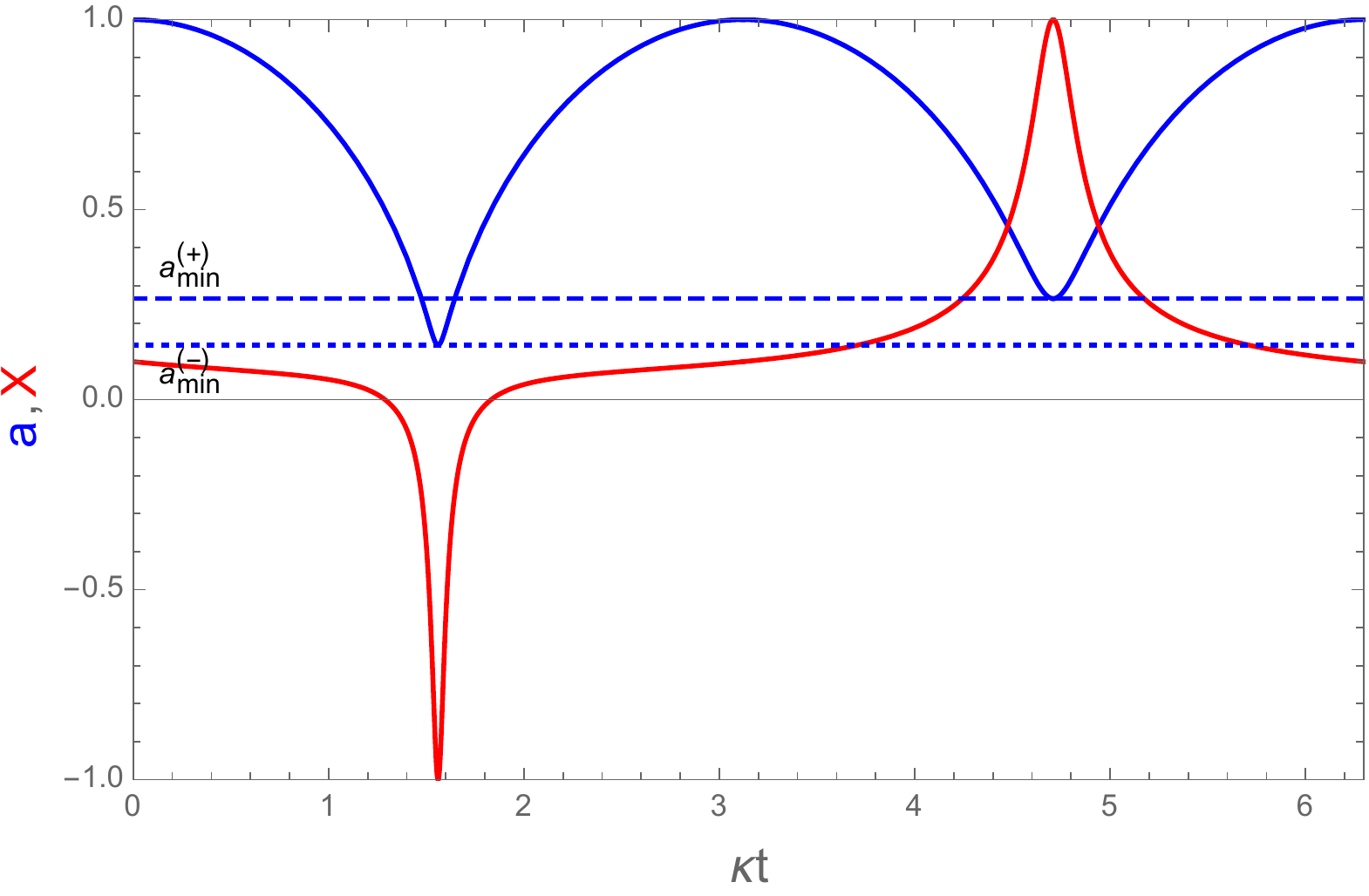}
   \caption{Evolution of $a$ (blue) $X$ (red). As expected from Eq.~(\ref{e.dyncyc}), the dynamics enjoys a cycle of two oscillations compared to $a_{\rm min}^{(+)}$ (dashed) and $a_{\rm min}^{(-)}$ (dotted). The two cycles are labeled by the sign of $X$, that is by $\varepsilon$.}
   \label{fig:evol} 
\end{figure}

The dynamics is completely described by Eqs.~(\ref{e.s1}-\ref{e.s2}). As anticipated from the analysis of Fig.~\ref{fig:psi1}, it is clear that the system undergoes oscillations. Starting from the initial state $(a_0=1,X_0=\lambda, \dd X/\dd\tau>0$), it will reach the minimum and bounce. But when it reaches $a_0$ again, the sign of the derivative of $X$ has been switched so that it starts another cycle with the same initial conditions but with $\dd X/\dd\tau<0$. Hence, we have a succession of cycles with a period of 2 phases related to the $(-1)^n$ in Eq.~(\ref{e.cyclen}), i.e.
\begin{eqnarray}\label{e.dyncyc}
\begin{pmatrix} a \\ X \end{pmatrix} : &&
\begin{pmatrix} a_0 =1 \\ X_0 =\lambda \end{pmatrix} \rightarrow
\begin{pmatrix} a_{\min}^{(-)} \\ -1 \end{pmatrix} \rightarrow
\begin{pmatrix} a_0 =1 \\ X_0 =\lambda \end{pmatrix} 
\rightarrow
\begin{pmatrix} a_{\min}^{(+)} \\ +1 \end{pmatrix} \rightarrow
\begin{pmatrix} a_0 =1 \\ X_0 =\lambda \end{pmatrix}.
\end{eqnarray}
The numerical integration of the system~(\ref{e.s1}-\ref{e.s2}) is depicted on Fig.~\ref{fig:evol} and compared to the minimal radii~(\ref{e.defamin}). It confirms that there is an alternance of cycles related to the change of sign of $X$  at the maximum expansion  in $a=1$. This tale of two bounces was first found numerically in  Refs.~\cite{Corichi:2011pg,Dupuy:2016upu}. However, to our knowledge, the full analytical derivation of the two minima appears for the first time here, as well as the explanation for the shift between the two cycles. 

\subsubsection{Cosmological implications}

The former description applies both to the dynamics of a collapsing star and to cosmology. It can be easily extended to consider matter with different equations of state, in particular to include a radiation fluid or a cosmological constant. We can rephrase the modifications of the Friedmann equations as the effects of an effective fluid with density $\rho_Q$ and pressure $P_Q$ that enter the standard Friedmann equations. It follows that
\begin{eqnarray}
\rho_Q + P_Q &=& -(\rho+P)\Psi_2(a), \\
\frac{8\pi G}{3}\rho_Q&=& -\left( \frac{8\pi G}{3}\rho - \frac{1}{R_c^2 a^2}\right)\Psi_1(a).
\end{eqnarray}
These expressions show that generically the effect of the loop regularization is to induce the action of an effective fluid with negative pressure. The conservation equation $\rho'+3{\cH}(\rho+P)=0$ gives the relation between $\Psi_2$ and $\Psi_1$ that generalizes Eq.~(\ref{e.xpsi2}) so that we end up with
\begin{eqnarray}
\rho_Q&=& \left(-\rho+\frac{3}{8\pi G R_c^2 a^2}\right)\Psi_1(a) \\
\rho_Q + P_Q &=&-\left[(\rho+P)-\frac{1}{4\pi G R_c^2 a^2} \right]\Psi_1(a) +\left( \frac{1}{3}\rho - \frac{1}{8\pi G R_c^2 a^2} \right) a \frac{\dd\Psi_1}{\dd a}.
\end{eqnarray}
Assuming that the standard matter enjoys a constant equation of state $w$, so that $\rho = \rho_0 a^{-3(1+w)}$, we get
\begin{eqnarray}
\rho_Q&=& -\rho_0 \left( a^{-3(1+w)}- \frac{3}{8\pi G R_c^2\rho_0 a^2}\right)\Psi_1(a) \\
\rho_Q + P_Q &=&-\rho_0\left[(1+w)a^{-3(1+w)}-\frac{2}{8\pi G R_c^2\rho_0 a^2} \right]\Psi_1(a) \nn \\
&&-\rho_0\left( -\frac{1}{3}a^{-3(1+w)} + \frac{1}{8\pi G R_c^2\rho_0 a^2} \right) a \frac{\dd\Psi_1}{\dd a}.
\end{eqnarray}
Setting $\kappa_0\equiv \frac{1}{8\pi G R_c^2\rho_0}$ we get
\begin{eqnarray}
\rho_Q&=& -\rho_0 \left( a^{-(1+3w)}- 3\kappa_0\right)\frac{\Psi_1}{a^2} \\
P_Q &=&-\rho_0\left[wa^{-(1+3w)}+ \kappa_0 \right]\frac{\Psi_1}{a^2} -\rho_0\left( -\frac{1}{3}a^{-(1+3w)} + \kappa_0\right) \frac{1}{a} \frac{\dd\Psi_1}{\dd a}.
\end{eqnarray}
These expressions can be used to determine the quantum corrections to the expansion history in a radiation universe or during inflation, a general phenomenological analysis that we postpone for now.

\section{Modeling a pulsating Planck star}\label{section4}

The dynamics of a bouncing homogeneous spacetime can be used to describe the interior of a collapsing star. We now turn to the construction of the Planck star model since the whole study of the UV-complete closed universe applies to the dynamics of the interior of such a star.  What remains to be derived are indeed the physical properties of the thin shell, namely its energy and pressure profiles and discuss the semi-classical limit of the model. Finally, one has to identify the sub-region of the parameters space for which the modelisation of the star is well-defined. 

\subsection{Quantum vs classical collapse}

As a first step, we present the classical Oppenheimer-Snyder model with a non-vanishing thin shell which should correspond to the classical limit of our quantum extension.

\subsubsection{The classical limit}

As discussed in the first section, the seminal OS model is constructed without assuming the presence of a thin-shell \cite{Oppenheimer:1939ue}. Hence, the present construction with $\Psi_1=0$ provides a classical extension of the initial OS model. Moreover, since $\Psi_1 =0$ corresponds to the limit 
\be
\lambda  = \frac{\tilde{\lambda}}{R_c}\rightarrow 0 \qquad \Rightarrow \qquad a^{\pm}_{\text{min}} =0
\ee
this model actually corresponds to the classical limit of our Planck star construction for which $a \in [0,1]$ such that the star can form a singularity. Notice that the classical limit, at fixed $\lambda$, corresponds to send the free parameter $R_c$ to larger values, and thus to consider compact objects with lower and lower density, as expected.

Let us first discuss the reduced mass relation. Working in the classical framework with $\Psi_1=0$, it follows that the discriminant~(\ref{e.det00}) is positive only if
\begin{equation}\label{e.cont2}
\sin\chi_s \leq \sin^3\chi_0.
\end{equation}
This provides a generalization of the seminal OS model, which corresponds to the case where the inequality is saturated. Solving the energy profile of the thin shell in this generalized classical OS model, one finds there are two roots for $\Sigma$ given by
\begin{eqnarray}
\label{GR}
\Sigma^{GR}_{\pm} &=&\frac{1}{a} \left[\cot{\chi_0} \pm \sqrt{\cot^2\chi_0+ \frac{1}{a}\left( \frac{\sin\chi_s}{\sin^3\chi_0}-1\right)} \right], \nonumber\\
\end{eqnarray}
which diverge both at the singularity $a=0$, as expected. The first branch $\Sigma^{GR}_{+}$ is always positive while the second branch can be negative near the singularity where the GR description is expected to break down. Interestingly, only the branch $\Sigma^{GR}_{-}$ vanishes in the OS case where $\sin{\chi_s} = \sin^3{\chi_0}$. Otherwise, the positive branch $\Sigma^{GR}_{+}$ has always a non vanishing energy carried by the shell even prior to collapse. 
Finally, the total energy of the shell, which scales as $\Sigma^{GR}_{\pm} a^2$ remains bounded for $a\in[0,1]$ in both branches. 

\subsubsection{Comparison of the classical and quantum dynamics}

\begin{figure}[h]
 \centering
   \includegraphics[scale= .52]{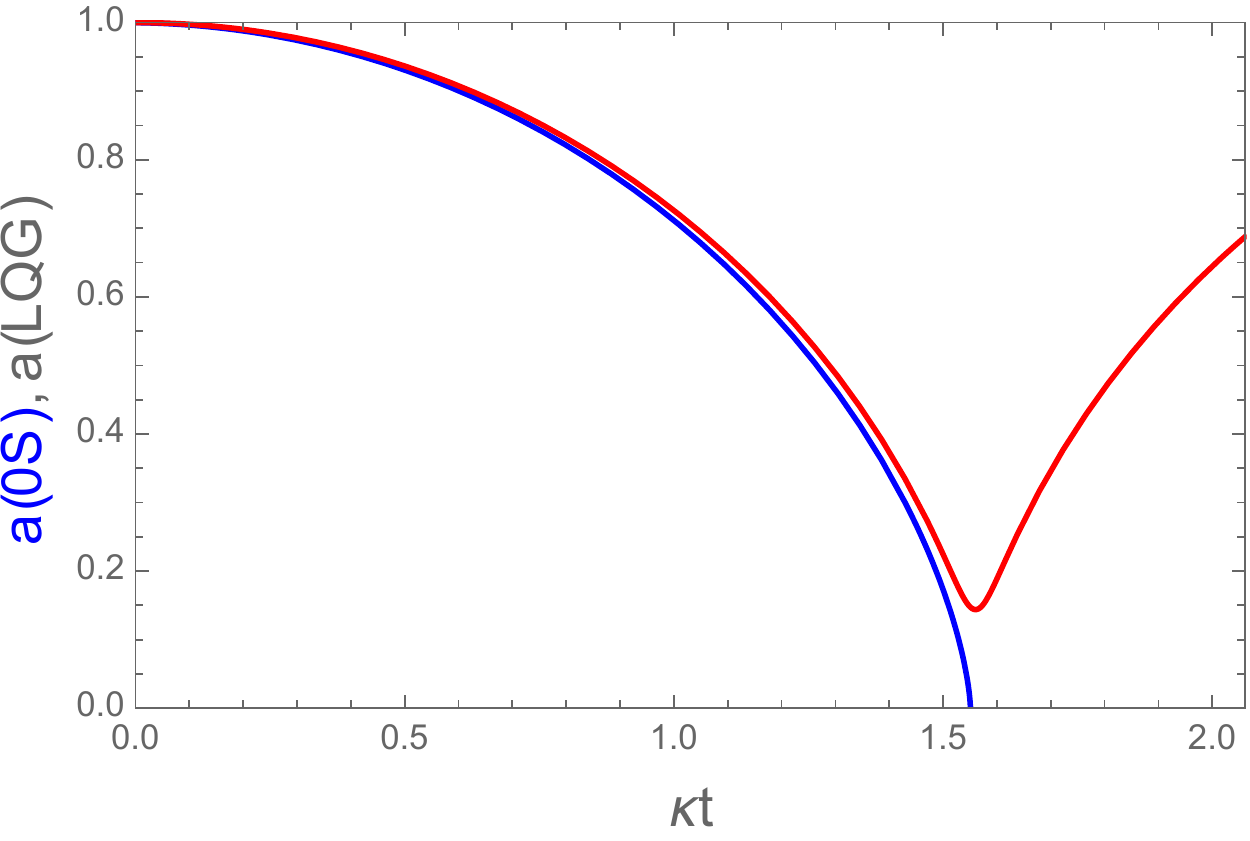}\hskip.2cm\includegraphics[scale= .53]{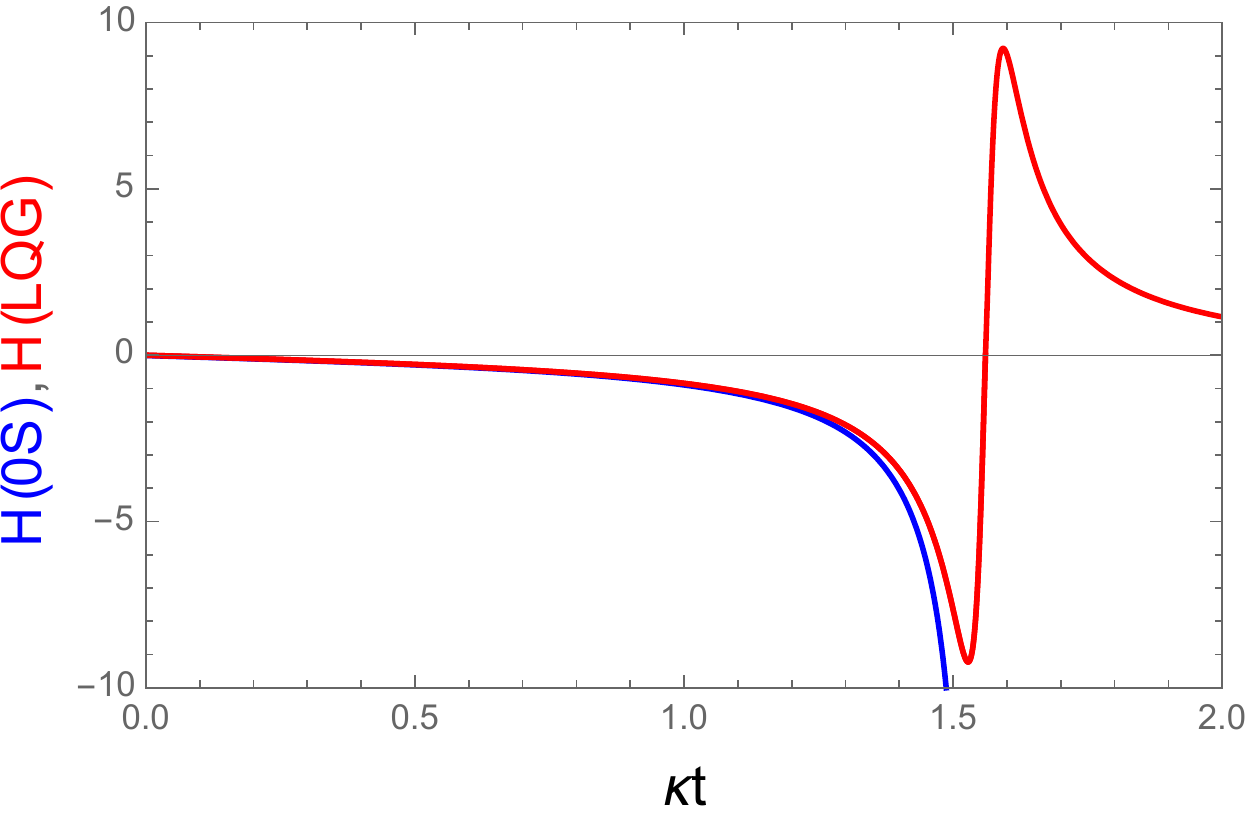}
   \caption{Comparison of the quantum induced bounce and the RG dynamics: (Left) Evolution of the scale factor for our Planck star (red) versus the classical GR collapse with thin shell (blue). (Right) Evolution of the Hubble rate for the Planck star (red) versus the classical GR collapse with a thin shell (blue).}
   \label{fig:LQC-RG} 
\end{figure}

Let us first check that the modified dynamics matches with the classical relativistic dynamics when quantum effects become negligible, namely far away from the bounce. As one can see from Fig.~\ref{fig:LQC-RG}, this is indeed the case. It confirms that our model has the correct classical limit far from the bounce and matches properly with the GR dynamics. Hence, the quantum corrections dominate only close to the bounce. 

\begin{figure}[h]
 \centering
   \includegraphics[scale= .5]{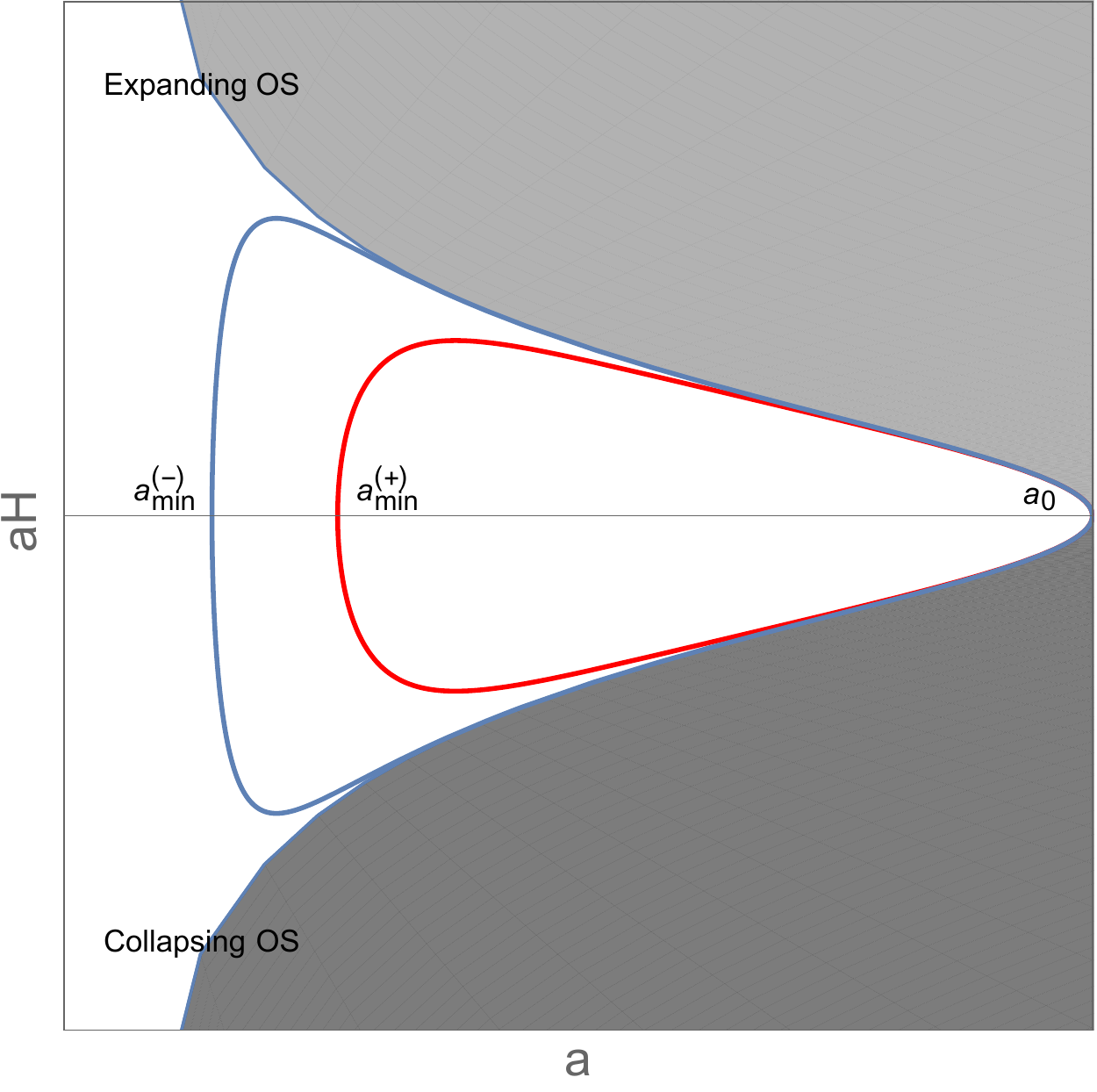} 
   \caption{Dynamics of the collapsing OS star (black) compared to its quantum version, which exhibits cycle of two bounces (red and blue).}
   \label{fig:phasespace} 
\end{figure}

The two dynamics can also be compared in the phase space $(a,\dot a)$ depicted  on  Fig.~\ref{fig:phasespace}. It clearly shows that the OS trajectory for the collapsing star (and is symmetric describing an expanding star) defines a boundary in phase space in which the cycles of bounces allowed by quantum corrections are enclosed. It clearly demonstrates how these corrections allow the dynamics to shift from the collapsing branch to the expanding branch with two possible bouncing minimal radii.
 
\newpage

\subsubsection{Properties of the effective thin shell}

Let us now discuss the properties of the shell in our effective quantum extension of the OS model. As in GR, there are two branches $\Sigma_{\pm}$ given by (\ref{e.defSig1}). The first one satisfies $\Sigma_+ > 0$ while the second one can have negative values. In order to allow for an exotic energy-momentum on the shell $\cT$, which shall be partially induced by quantum effect, it is therefore natural to focus on the second branch $\Sigma_{-}$. It reads explicitly
\begin{align}\label{e.det00}
\Sigma_{-} = \frac{1}{a} \left\{ \cot{\chi_0} - \sqrt{\cot^2\chi_0  - \frac{1}{a}\left[ \frac{\sin\chi_s}{\sin^3\chi_0} -1 + (1-a)\Psi_1 (a) \right]}\right\} 
\end{align}
with $a \in [1,a^{\pm}_{\text{min}}]$ and $\Psi_1(a)$ given by (\ref{e.Psi1deA}). When $a=1$, the quantum corrections vanish and do not affect the classical limit which reproduces exactly the GR expression (\ref{GR}). Therefore, the behavior of the shell is consistent with the expected classical limit and with the expectation that quantum effects are triggered only near the bounce. Finally, using the point of view explained in Section~\ref{sub2c}, the thin shell energy-momentum tensor $\cS_{ab}$ satisfies automatically some continuity equations which are a direct consequence of the Israel-Darmois junction conditions together with the scalar and momentum constraint of GR imposed in $\cT$. See \cite{Israel:1966rt,Barrabes:1991ng} for details. Recall that these continuity equations are valid provided the exterior and interior geometries are understood as GR solutions associated to some exotic energy-momentum tensor, a point of view which is always available due to the freedom to view quantum corrections either on the geometry or matter sector.

\subsection{Orders of magnitude}

\label{om}

Finally, let us discuss the constraints on the order of magnitude associated to such a Planck star. The compact object described by our model is parametrized by five free parameters: $\left(\rho_{\rm min}, R_{\text{max}}, M\right)$ which correspond to the physical parameters of the star, respectively to its initial density, its maximal radius and its mass together with the parameters of the theory, i.e. $( \tilde{\lambda}, \gamma)$ which encode the UV cut-off of the quantum theory and a freedom in defining the Ashteklar's variables. The BI parameter $\gamma$ is mostly irrelavent and of order unity. The four remaining parameters need to satisfy
\be\label{conditioncruciale}
R_s  \leqslant  \left(\frac{\gamma\tilde{\lambda}}{R_c}\right)^{2/3} R_{\rm max}\qquad \text{and} \qquad \tilde\lambda<R_s < R_{\text{max}} < R_c
\ee
The first condition is indeed the constraint~(\ref{kkin}) in which we used the fact that for small $\lambda$, $a_{\rm min}\sim \gamma^{2/3}\lambda^{2/3}$. The first inequality of the second condition is added to ensure that the physical description is valid. Moreover, in order for our effective description to remain valid, the mass of the compact object has to be well above the Planck mass, i.e. $\text{M}/ \text{M}_{\text{Planck}} \gg 1$. 

\begin{figure}[h]
 \centering
   \includegraphics[scale= .5]{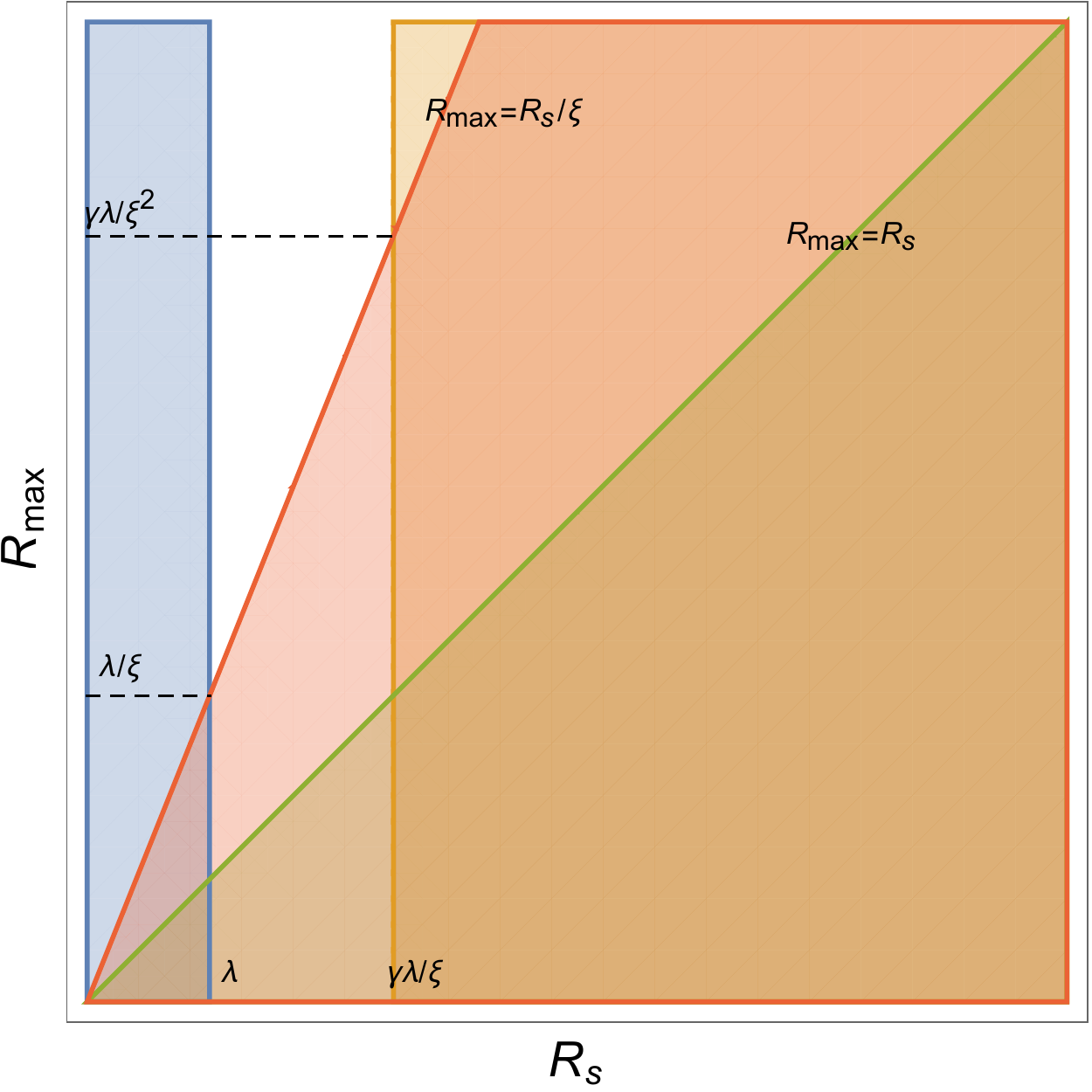}
   \caption{The parameter space of stars $(R_s,R_{\rm max})$ that are allowed given the parameters of the theory $(\tilde\lambda,\xi)$. Note that $\tilde\lambda$ fixes all the scales of the model.}
   \label{fig:modelOK} 
\end{figure}

In order to capture the domain of applicability imposed by the above consistency conditions, it is useful to introduce the quantity $\xi$ defined by
\be\label{defxi}
\xi^2\equiv \frac{4\pi}{3}\rho_{\rm min}\frac{R_{\rm max}^3}{M},
\ee
so that a star without shell in GR will have $\xi=1$. Therefore, the parameter $\xi$ encodes the mass spliting between the bulk interior region and the shell. For small $\xi$, most of the mass of the compact object is carried by the thin shell, and the bulk is very diluted.  Using this parameter $\xi$, it follows that
\be
R_c = \xi \frac{R_{\rm max}^{3/2}}{R_s^{1/2}}.
\ee
Hence, Eq.~(\ref{conditioncruciale}) rewrites as
\be\label{conditioncruciale2}
R_s  \leqslant  \frac{\gamma\tilde{\lambda}}{\xi}
\qquad \text{and} \qquad 
\tilde\lambda<R_s < R_{\text{max}} < \xi\frac{R_{\rm max}^{3/2}}{R_s^{1/2}}.
\ee
We first conclude that independently of the radius of the star, its mass is constrained to be in the range
\be\label{conditionRs}
\tilde\lambda \leqslant  R_s  \leqslant  \frac{\gamma\tilde{\lambda}}{\xi},
\ee
from which we conclude that necessarily one need $\xi<1$. Then the second set of constraints reduces to the boundary $R_s>\tilde\lambda$, $R_{\rm max}>R_s/\xi$ and $R_{\rm max}>R_s$, which is irrelevant when $\xi<1$! The space of allowed models is summarized on Fig.~\ref{fig:modelOK}. Note that $\tilde\lambda$ fixes all the scales of the problem and Eq.~(\ref{conditionRs}) implies that the mass of the star we want to describe imposes strong bounds on $\xi$ if we want to have $\tilde\lambda\sim{\cal O}(\ell_P)$ and that its minimum radius is larger than $(\gamma/\xi)^{2/3}\tilde\lambda$ and the maximum density inside the star is smaller than $\rho_{\rm min}(\xi/\gamma)^2(R_{\rm max}/\tilde\lambda)^3$. Introducing $\hat\lambda\equiv\tilde\lambda/\ell_P$ we have
\be
\hat \lambda \leqslant  1.8\times10^{38}\frac{M}{M_\odot}  \leqslant  \frac{\hat{\lambda}}{\xi},
\ee
where we have set $\gamma=1$. Note that this implies that
\be
\sin\chi_s = \frac{1}{\xi}\left(\frac{R_s}{R_{\rm max}}\right)^{3/2}, 
\qquad
\sin\chi_0 = \frac{1}{\xi}\left(\frac{R_s}{R_{\rm max}}\right)^{1/2}, 
\ee
so that $\sin\chi_s=\xi^2\sin^3\chi_0$.

Let us take two examples. First consider a Planck relic with $M = 10^{12} \text{kg}$ and $R_{\text{max}} \simeq 10^{-14} \; \text{m}$ which mimics the order of magnitude considered initially in Ref. \cite{Rovelli:2014cta}. It shall satisfy $\hat \lambda \leqslant 10^{20} \leqslant \hat{\lambda}/\xi$ so that $\xi$ has to be smaller than $10^{-20}$ and the smallest star that can be described is about $R_{\rm max}\sim10^{20}\ell_P\sim 10^{-15}$~m. Now, if one assumes $M=1$~kg then $\hat \lambda \leqslant 10^{8} \leqslant \hat{\lambda}/\xi$ so that $\xi<10^{-8}$ and one can have $R_{\rm max}\sim10^{8}\ell_P$. For Stellar mass objects then $M/M_\odot\sim1$ so that one would need $\xi<10^{-38}$. While mathematically possible, this will lead to stars with a density $10^{-38}$ smaller than the Solar density, hence smaller than $10^{-35}$~kg/m$^3$, which is indeed unrealistic. Notice that since $\xi$ has to be very small for our model to make sense, all the objects described by this model will have a small portion of their mass carried by the bulk interior while most of their mass will be carried by thin shell. This is a consequence of the consistency condition (\ref{conditioncruciale2}).

These orders of magnitude show that one can easily set $\tilde\lambda$ of the order of the Planck scale and describe Planckian relic with the present model, but stellar massive compact objects cannot be modeled in a realistic manner within this construction. In turn, this restriction on the range of applicability of the model ensures that while the bounce takes place above or at the Schwarzschild radius, it still occurs in a high curvature regime as these Planckian relics exhibit already very high curvature at the horizon scale. Finally, notice that within the range of masses discussed in the above examples, the Planckian relics satisfy the conditions $\text{M}/ \text{M}_{\text{Planck}} \gg 1$, which ensures the validity of the effective equations even for such extremely small mass and small size objects. See Ref. \cite{Rovelli:2013zaa} for a discussion on this point. 
 
The present model is obviously highly simplified and the objects described by this construction are not realistic. However, this exercise allows one to investigate the different challenges one has to face when constructing a mathematically consistent model of bouncing compact object based on the most simple scenario of matter collapse instead of eternal vacuum black hole interior.

\subsection{Comparison with other models}

As a final step, let us compare our model to previous constructions. The main outcome of the present construction relies in the fact that the bounce occurs at a radius above or at the Schwarzschild radius. As such, the quantum corrections are not confined to the deep interior, but they are triggered at least at the horizon scale, or even on larger scales depending on the choice of the parameter $\lambda$ (ad therefore $R_c$). This provides a major difference with current black-to-white hole bounce models recently discussed in the LQC literature, where quantum effects are modeled such that they become dominant only in the deep interior region. A lesson of the present construction is that depending on the assumptions of the model, the scale at which the quantum effects become non-negligible shall not be set a priori, but should be dictated by the internal consistency of the construction. In the present case where we consider dynamical matter collapse, consistency imposes that quantum gravity effects are actually relevant above and at the Schwarzschild radius. This appears as a kinematical condition in the thin shell construction. A major lesson from this model, and in particular of the constraint (\ref{KEY}), is that if a consistent model of black-to-white hole bounce has to be build from a matter collapse scenario, quantum gravity effects can be confined to the deep interior only if this model allows the formation of an inner horizon. From this point if view, the above result suggests that the recent models discussed in \cite{Ashtekar:2018cay, Bodendorfer:2019xbp}, which focus on the vacuum Schwarzschild interior and do not exhibit an inner horizon, are missing a crucial ingredient. If matter is included, such additional structure turns out to be crucial to keep quantum gravity effects confined in the deep interior. A concrete realization of this scenario has been presented in Ref. \cite{next}. 

However, notice that in the present construction, even if the bounce occurs above the Schwarzschild radius, it does not imply that quantum gravity effects are triggered at low curvature. As we shall see in Section~\ref{om}, the compact objects which can be consistently described by this model are constrained by the conditions (\ref{conditioncruciale}). This restriction of the parameters of the model ensures that only very small mass and small size objects, such as Planckian relics, can be described while macroscopic stellar objects are excluded. As it turns out, such Planckian relics exhibit already huge curvature even outside their Schwarzschild radius, and therefore, the bounce can consistently occur at high curvature regime for such objects, even outside the would be horizon.

As a final remark, let us also point that the present model use a regularization of the interior geometry based on the $\bar{\mu}$-scheme commonly used in LQC. This is another difference with current polymer constructions of regular eternal black holes interior. Indeed, the recent constructions \cite{Ashtekar:2018cay, Bodendorfer:2019xbp} implement a modified $\mu_{\circ}$-scheme regularization. However, such choice is somehow not consistent with the standard improved dynamics used in the cosmological framework, and it would appear more natural to adopt the same strategy both for the loop regularization of cosmological and black hole backgrounds. From this point of view, our model provides such a minimal construction using the standard LQC techniques based on the $\bar{\mu}$-scheme regularization.

\section{Conclusion}

Building on the effective construction presented in our companion article \cite{paperUS}, this work has presented a new physical model for a bouncing compact object based on a quantum extension of the seminal OS model. This simple model of gravitational collapse, based on the thin-shell formalism, turns out to provide an ideal framework to bridge bouncing cosmological models to the description of bouncing compact objects. Let us summarize its main ingredients and its major outcomes.

First, the singularity resolution inside the interior of the star is obtained by replacing the classical closed FL universe by its LQC version based on the connection regularization~\cite{Corichi:2011pg, Dupuy:2016upu}. This loop regularization provides a concrete proposal for how the would-be singularity is replaced by a singularity free effective quantum geometry. 

Then, our construction includes the matter sector. This allows us to go beyond current polymer approaches which largely focus solely on the vacuum spherically symmetric gravity phase space to discuss effective black-to-white hole bounce\footnote{See however Ref.~\cite{Joe:2014tca} for a study including matter in the Kawtowschi-Sachs model and Refs.~\cite{Gambini:2014qga, BenAchour:2016brs, BenAchour:2017jof, Bojowald:2019fkv} for few investigations regarding covariant matter plus gravity systems in inhomogeneous polymer models.}. Moreover, our model provides also an alternative to matter collapse compared to the firework model, where matter is encoded in self-gravitating null shell. In our construction, even if highly idealized, the dust field provides a simpler and more physical matter content, and allows one to work with a time-like thin shell. This difference is crucial and is at the heart of our no-go result on the formation of trapped region. Indeed, with a null shell, one would not derive the expression (\ref{keyrel}) for the lapse at the surface of the star. This explains the radical difference between our conclusions and those of the firework model. 

The time-like thin-shell is crucial for the internal consistency of our approach. Indeed, it encodes part of the quantum effects and leads to a generalized version of the OS mass relation. The key constraint,  given by Eq. (\ref{e.massrel}), ensures the consistent matching of the conserved quantities of the exterior and interior geometries. From the general solution of the junction conditions presented in Ref. \cite{paperUS}, we have discussed the behavior of the energy and pressure profile of the thin shell associated to a LQC bouncing interior geometry. As expected, quantum corrections become large near the bounce, and vanish far away for it. The classical limit of the model, which consists in the OS model with a non-vanishing thin shell, is well approximated at large radius.

The surprising outcome of our construction is that replacing the singular classical interior geometry with a bouncing geometry prevents from forming a trapped region. Hence, in such a UV complete gravitational collapse model, there is no formation of black hole. This provides a major deviation from the classical OS description of the collapse. We emphasize that this conclusion descends from i) demanding the continuity of the induced metric across the time-like thin shell and ii) modeling the exterior geometry with the Schwarzschild geometry which exhibits a single horizon. Relaxing this assumption and using an exterior geometry with an outer and inner horizon would allow the collapsing matter to form a trapped region and bounce inside the inner horizon. The present construction suggests therefore that the formation of an inner horizon is needed to build consistent black-to-white hole models based on matter collapse in which quantum gravity effects remain confined to the deep interior, in a high curvature regime.

Let us now summarize the phenomenology of our model. Since, the interior geometry enjoys both a UV cut-off, i.e $\tilde{\lambda}$ descending from the loop regularization, and an IR-cut-off $R_c$ which encodes the energy density of the star prior to collapse. The compact object follows therefore cycles of collapse and expansion, with two consecutive distinct bounces. A novelty of the present construction is that the IR cut-off $R_c$ is now a free parameter of the model, contrary to the classical OS model where it is fixed once the mass and maximal radius are chosen, i.e once $\left( \chi_s, \chi_0\right)$ are fixed. This freedom allows to explore new compact objects, which are much denser than the one described in GR, although not forming black holes. As it turns out, the energy scale at which quantum gravity becomes non-negligible is encoded in the ratio $\tilde{\lambda}/R_c$. Hence, at fixed $\tilde{\lambda}$, shifting the minimal energy of the compact object encoded in $R_c$ allows to shift the scale at which quantum gravity becomes dominant. Finally, we have shown that the no-go result associated to this model, which translates into the constraint (\ref{conditioncruciale}) onto the parameters, implies that this construction can be applied only to compact objects like Planckian relics but not to standard macroscopic stellar objects. This restriction of the range of applicability reconciles the no-go result presented in our companion paper \cite{paperUS} with the expectation that a quantum bounce, as the one considered here, shall occur only in a high curvature regime. Moreover, it is worth pointing that one can exhibit examples of such Planckian relics which satisfy the condition $\text{M}/ \text{M}_{\text{Planck}} \gg 1$ such that the effective equations used in the present construction are still valid. This shows that, despite the rather surprising no-go result imposing the bounce above or at the Schwarzschild radius, the present highly idealized model can still be used to consistently model some range of Planckian relics. 

In conclusion, this model provides a highly idealized model of bouncing compact object within the LQC framework which allows one to realize several crucial ideas introduced in the initial Planck star model in Ref. \cite{Rovelli:2014cta}. This construction should serve as a minimal set-up for further explorations regarding the modelization of bouncing compact objects using LQC techniques. In particular, it would be interesting to compute the characteristic time of bounce associated to such Planckian relics and discuss its scaling w.r.t. the mass of the object. Additionally, it would be interesting to investigate the stability of such object, as an initial inhomogeneity would grow during the oscillation and the compact object would most probably follow only a finite number of cycles before being out of the range of validity of the present model. Finally, several simplifying assumptions of the present construction might be generalized to consider black-to-white hole bounce for macroscopic stellar objects, such as introducing an inner horizon structure in the exterior geometry. We shall present some results in this direction in future works \cite{next}.

 \section*{Acknowledgements}
The work of JBA was supported by Japan Society for the Promotion of Science Grants-in-Aid for Scientific Research No. 17H02890. The authors thank Norbert Bodendorfer and Edward Wilson-Ewing for helpful suggestions on a previous version of this draft.

\appendix

 \section{Scaling properties of the physical quantities under rescaling of $R_c$}

\label{scaling}

This Appendix investigates the scaling properties of the various variables of the model when we rescale the free parameter $R_c$. First of all, we recall that this parameter, which encodes the spatial curvature of the closed universe, can be safely set to unity without loss of generality when dealing with the cosmological solution alone. However, it is not true anymore when one glues this interior geometry to an exterior one since then it plays a key role.  As explained, it is related to the minimal energy of the compact object through
\be
\rho_{\text{min}} = \frac{3}{8\pi G R^2_c}
\ee
We have pointed in Section~\ref{main} that because of the thin shell, this minimal energy is no more determined by the mass and size $(R_s, R_{\text{max}})$  of the compact object prior to collapse as in the OS model. Now, one is free to split the total energy of the object between the interior bulk region and the thin shell. This freedom allows one to play with the density of the bulk interior region and this freedom is reflected in the free parameter $\rho_{\text{min}}$, or equivalently $R_c$. It is therefore interesting to investigate how each quantity behaves under a rescaling of $R_c$. Consider the transformation
\be
R_c \rightarrow \hbox{e}^{\epsilon} R_c \;, \qquad \text{or equivalently} \qquad \rho_{\text{min}} \rightarrow \hbox{e}^{-2\epsilon} \rho_{\text{min}} .
\ee
where $\epsilon$ is a constant positive real number and let fix once and for all the UV cut-off $\tilde{\lambda}$. Then, the different scales of the problem transform as
\begin{align}
R_{\text{max}} & \rightarrow \hbox{e}^{\epsilon} R_{\text{max}} \\
\label{massrescaling}
R_{s} & \rightarrow \hbox{e}^{\epsilon} R_{s} \\
\lambda & \rightarrow \hbox{e}^{-\epsilon} \lambda.
\end{align}
Therefore, the total mass of the object $R_s= 2GM$ gets rescaled as well as the effective UV cut-off $\lambda$ of the model. Then, using the different definitions of the dynamical physical quantities entering in the model, we have
\begin{align}
R  \rightarrow \hbox{e}^{\epsilon} R, \quad
\rho  \rightarrow \hbox{e}^{-2\epsilon} \rho, \quad
 \sigma  \rightarrow \hbox{e}^{-\epsilon} \sigma 
\end{align}
while the rescaled surface energy $\Sigma = 8\pi G R_c \sigma$ and the quantum correction $\Psi_1$ remained unaffected
\begin{align}
 \Sigma  \rightarrow \Sigma, \qquad  \Psi  \rightarrow \Psi_1.
\end{align}
Notice that the quantum correction $\Psi_1$ in (\ref{qcor}) is defined after a rescaling of the Hubble parameter $b \rightarrow R_c b$ which cancels the rescaling of the effective UV cut-off $\lambda$ entering in its argument. Equipped with these transformations, we can now check the transformation property of the junction condition. After such rescaling, it is direct to show that 
\begin{align}
M & \rightarrow  \frac{4\pi}{3} \left( \hbox{e}^{-2\epsilon} \rho\right) \left( \hbox{e}^{3\epsilon} R^3\right) - \frac{\left( \hbox{e}^{3\epsilon}R^3\right) }{2G \left( \hbox{e}^{2\epsilon} R^2_c\right)} \left\{ \Sigma - 2 \frac{\left( \hbox{e}^{\epsilon} R_c\right)}{\left( \hbox{e}^{\epsilon} R\right)} \cos{\chi_0} \Sigma + \frac{1-a}{a^3} \Psi_1\right\} \\
& = \hbox{e}^{\epsilon} \left[ \frac{4\pi}{3} \rho R^3 - \frac{R^3 }{2G  R^2_c } \left\{ \Sigma -  \frac{ 2R_c}{R} \cos{\chi_0} \Sigma + \frac{1-a}{a^3} \Psi_1\right\} \right]\\
 &= \hbox{e}^{\epsilon} M,
\end{align}
which shows the consistency of the quantum corrected junction condition with the expected scaling of the mass of the object (\ref{massrescaling}), as well as its invariance under this transformation. This indicates that a rescaling of the minimal energy density of the compact object prior to collapse at fixed UV cut-off $\tilde{\lambda}$ implies a rescaling of the total mass of the object. Now, it is well known that in spherically symmetric General Relativity, the total mass is canonically conjugated to the asymptotic Killing time; see Ref.~\cite{Kuchar:1994zk} for details. Therefore, if one requires that the symplectic 2-form associated to the system be invariant under such rescaling, one has to perform an additional rescaling of the asymptotic Killing time variable.

\section{On the analytic resolution of the interior dynamics}

\label{B}

Consider the equations dictating the interior bouncing dynamics presented in Section~(\ref{RFDYN}). Using the auxiliary fields $(u,X)$ defined in Eq.~(\ref{e.cvar}), the dynamics is given by the system~(\ref{e.s1}-\ref{e.s2}) together with the constraint~(\ref{e.c00}). Now, let us introduce the new time coordinate
\be
\dd \tilde{\tau} = \cos{(\lambda b)} \dd \tau \;.
\ee
Notice that the lapse $\dd \tilde{\tau} / \dd \tau$ now changes sign during the dynamics.  With respect to this new time and upon using the scalar constraint, one can turn the  system~(\ref{e.s1}-\ref{e.s2})  into
\begin{eqnarray}
\frac{\dd u_{\pm}}{\dd \tilde{\tau}} \;  &\widehat{=}& \mp u_{\pm}^2 \sqrt{u_{\pm}-1}\label{e.XX1}\\
\frac{\dd X_{\pm}}{\dd \tilde{\tau}} \;  &\widehat{=}& -\lambda\gamma u_{\pm}\left\lbrace 
\frac{3}{2}u_{\pm}^2 - u_{\pm} \pm\frac{u_{\pm}\sqrt{u_{\pm}-1}}{\gamma}\label{e.XX0}
\right\rbrace
\end{eqnarray}
where $ \widehat{=}$ means that we have used the scalar constraint (\ref{e.c00}) and $u_{\pm}$ corresponds to the ambiguity in the choice of sign when taking the square root of the scalar constraint as
\be
\label{e.c01}
X_{\pm}-\lambda u_{\pm}={\pm}\gamma\lambda u_{\pm}\sqrt{u_{\pm}-1},
\ee
that was used to simplify the r.h.s. of Eq.~(\ref{e.s2}). The first equation is easily integrated to give
\begin{align}
\label{e.c02}
 \frac{\sqrt{u_{\pm} -1}}{u_{\pm}} + \arctan\sqrt{u_{\pm} -1} = \mp(\tilde{\tau}- c),
\end{align}
where $c$ is a constant of integration while $X_\pm$ is given, from Eq.~(\ref{e.c01}), by
\be
X_{\pm}=\lambda u_{\pm}\left[1 {\pm}\gamma\sqrt{u_{\pm}-1}\right],
\ee
and one can check that Eq.~(\ref{e.XX0}) is satisfied. Unfortunately, there is no obvious way to invert
expression~(\ref{e.c02}) to get $u_{\pm}(\tau)$. This justifies the numerical integration of the interior dynamics presented in Section~(\ref{interiornum}).


\begin{thebibliography}{ab}

 
\bibitem{Rovelli:2014cta}
C.~Rovelli and F.~Vidotto, ``{Planck stars},'' Int. J. Mod. Phys. {\bf D23}
  (2014), no.~12, 1442026,
\href{http://arXiv.org/abs/1401.6562}{{\texttt{arXiv:1401.6562}}}.

\bibitem{Ashtekar:2011ni}
A.~Ashtekar and P.~Singh, ``{Loop Quantum Cosmology: A Status Report},'' Class.
  Quant. Grav. {\bf 28} (2011) 213001,
\href{http://arXiv.org/abs/1108.0893}{{\texttt{arXiv:1108.0893}}}.

\bibitem{Barrau:2014hda}
A.~Barrau and C.~Rovelli, ``{Planck star phenomenology},'' Phys. Lett. {\bf
  B739} (2014) 405--409,
\href{http://arXiv.org/abs/1404.5821}{{\texttt{arXiv:1404.5821}}}.

\bibitem{Barrau:2015uca}
A.~Barrau, B.~Bolliet, F.~Vidotto, and C.~Weimer, ``{Phenomenology of bouncing
  black holes in quantum gravity: a closer look},'' JCAP {\bf 1602} (2016),
  no.~02, 022,
\href{http://arXiv.org/abs/1507.05424}{{\texttt{arXiv:1507.05424}}}.

\bibitem{Barrau:2016fcg}
A.~Barrau, B.~Bolliet, M.~Schutten, and F.~Vidotto, ``{Bouncing black holes in
  quantum gravity and the Fermi gamma-ray excess},'' Phys. Lett. {\bf B772}
  (2017) 58--62,
\href{http://arXiv.org/abs/1606.08031}{{\texttt{arXiv:1606.08031}}}.

\bibitem{Rovelli:2017zoa}
C.~Rovelli, ``{Planck stars as observational probes of quantum gravity},'' Nat.
  Astron. {\bf 1} (2017) 0065,
\href{http://arXiv.org/abs/1708.01789}{{\texttt{arXiv:1708.01789}}}.

\bibitem{Barrau:2018kyv}
A.~Barrau, F.~Moulin, and K.~Martineau, ``{Fast radio bursts and the stochastic
  lifetime of black holes in quantum gravity},'' Phys. Rev. {\bf D97} (2018),
  no.~6, 066019,
\href{http://arXiv.org/abs/1801.03841}{{\texttt{arXiv:1801.03841}}}.

\bibitem{Haggard:2014rza}
H.~M. Haggard and C.~Rovelli, ``{Quantum-gravity effects outside the horizon
  spark black to white hole tunneling},'' Phys. Rev. {\bf D92} (2015), no.~10,
  104020,
\href{http://arXiv.org/abs/1407.0989}{{\texttt{arXiv:1407.0989}}}.

\bibitem{DeLorenzo:2015gtx}
T.~De~Lorenzo and A.~Perez, ``{Improved Black Hole Fireworks: Asymmetric
  Black-Hole-to-White-Hole Tunneling Scenario},'' Phys. Rev. {\bf D93} (2016),
  no.~12, 124018,
\href{http://arXiv.org/abs/1512.04566}{{\texttt{arXiv:1512.04566}}}.

\bibitem{Brahma:2018cgr} 
  S.~Brahma and D.~h.~Yeom,
  ``{Effective black-to-white hole bounces: The cost of surgery,}''
  Class.\ Quant.\ Grav.\  {\bf 35}, no. 20, 205007 (2018)
  \href{http://arXiv.org/abs/1804.02821}{{\texttt{arXiv:1804.02821}}}.


\bibitem{Rovelli:2018cbg}
C.~Rovelli and P.~Martin-Dussaud, ``{Interior metric and ray-tracing map in the
  firework black-to-white hole transition},'' Class. Quant. Grav. {\bf 35}
  (2018), no.~14, 147002,
\href{http://arXiv.org/abs/1803.06330}{{\texttt{arXiv:1803.06330}}}.

\bibitem{DAmbrosio:2018wgv} 
  F.~D'Ambrosio and C.~Rovelli,
  ``{How information crosses Schwarzschild’s central singularity,}''
  Class.\ Quant.\ Grav.\  {\bf 35}, no. 21, 215010 (2018)
  \href{http://arXiv.org/abs/1803.05015}{{\texttt{arXiv:1803.05015}}}.
  
       \bibitem{Barcelo:2014cla}
C.~Barcelo, R.~Carballo-Rubio, L.~J. Garay, and G.~Jannes, ``{The lifetime
  problem of evaporating black holes: mutiny or resignation},'' Class. Quant.
  Grav. {\bf 32} (2015), no.~3, 035012,
\href{http://arXiv.org/abs/1409.1501}{{\texttt{arXiv:1409.1501}}}.

\bibitem{Barcelo:2015uff}
C.~Barceló, R.~Carballo-Rubio, and L.~J. Garay, ``{Black holes turn white
  fast, otherwise stay black: no half measures},'' JHEP {\bf 01} (2016) 157,
\href{http://arXiv.org/abs/1511.00633}{{\texttt{arXiv:1511.00633}}}.

\bibitem{Barcelo:2016hgb}
C.~Barceló, R.~Carballo-Rubio, and L.~J. Garay, ``{Exponential fading to white
  of black holes in quantum gravity},'' Class. Quant. Grav. {\bf 34} (2017),
  no.~10, 105007,
\href{http://arXiv.org/abs/1607.03480}{{\texttt{arXiv:1607.03480}}}.


\bibitem{Bianchi:2018mml}
E.~Bianchi, M.~Christodoulou, F.~D'Ambrosio, H.~M. Haggard, and C.~Rovelli,
  ``{White Holes as Remnants: A Surprising Scenario for the End of a Black
  Hole},'' Class. Quant. Grav. {\bf 35} (2018), no.~22, 225003,
\href{http://arXiv.org/abs/1802.04264}{{\texttt{arXiv:1802.04264}}}.

\bibitem{Rovelli:2018hba} 
  C.~Rovelli and F.~Vidotto,
  ``{Pre-Big-Bang Black-Hole Remnants and Past Low Entropy,}''
  Universe {\bf 4}, no. 11, 129 (2018)
  \href{http://arXiv.org/abs/1805.03224}{{\texttt{arXiv:1805.03224}}}.
  
  \bibitem{Rovelli:2018okm} 
  C.~Rovelli and F.~Vidotto,
  ``{Small black/white hole stability and dark matter,}''
  Universe {\bf 4}, no. 11, 127 (2018)
   \href{http://arXiv.org/abs/1805.03872}{{\texttt{arXiv:1805.03872}}}.
  
  \bibitem{Martin-Dussaud:2019wqc} 
  P.~Martin-Dussaud and C.~Rovelli,
  ``{Evaporating black-to-white hole,}''
  Class.\ Quant.\ Grav.\  {\bf 36}, no. 24, 245002 (2019)
     \href{http://arXiv.org/abs/1905.07251}{{\texttt{arXiv:1905.07251}}}.
     

     
     \bibitem{Kawai:2013mda}  
  H.~Kawai, Y.~Matsuo and Y.~Yokokura,
  ``{A Self-consistent Model of the Black Hole Evaporation,}''
  Int.\ J.\ Mod.\ Phys.\ A {\bf 28}, 1350050 (2013)
   \href{http://arXiv.org/abs/1302.4733}{{\texttt{arXiv:1302.4733}}}.
   
   \bibitem{Baccetti:2018qrp} 
  V.~Baccetti, S.~Murk and D.~R.~Terno,
  ``{Black hole evaporation and semiclassical thin shell collapse,}''
  Phys.\ Rev.\ D {\bf 100}, no. 6, 064054 (2019)
     \href{http://arXiv.org/abs/1812.07727}{{\texttt{arXiv:1812.07727}}}.
   
   \bibitem{Ho:2018fwq} 
  P.~M.~Ho, H.~Kawai, Y.~Matsuo and Y.~Yokokura,
  ``{Back Reaction of 4D Conformal Fields on Static Geometry,}''
  JHEP {\bf 1811}, 056 (2018)
   \href{http://arXiv.org/abs/1807.11352}{{\texttt{arXiv:1807.11352}}}.
  
  \bibitem{Ho:2019pjr} 
  P.~M.~Ho, Y.~Matsuo and Y.~Yokokura,
  ``{An Analytic Description of Semi-Classical Black-Hole Geometry,}''
    \href{http://arXiv.org/abs/1912.12855}{{\texttt{arXiv:1912.12855}}}.
  
  
  \bibitem{Bolokhov:2018rsa} 
  S.~V.~Bolokhov, K.~A.~Bronnikov and M.~V.~Skvortsova,
  ``{The Schwarzschild singularity: a semiclassical bounce?,}''
  Grav.\ Cosmol.\  {\bf 24}, no. 4, 315 (2018)
    \href{http://arXiv.org/abs/1808.03717}{{\texttt{arXiv:1808.03717}}}.
    
    \bibitem{Ashtekar:2005qt}
A.~Ashtekar and M.~Bojowald, ``{Quantum geometry and the Schwarzschild
  singularity},'' Class. Quant. Grav. {\bf 23} (2006) 391--411,
\href{http://arXiv.org/abs/gr-qc/0509075}{{\texttt{arXiv:gr-qc/0509075}}}.

\bibitem{Modesto:2006mx}
L.~Modesto, ``{Black hole interior from loop quantum gravity},'' Adv. High
  Energy Phys. {\bf 2008} (2008) 459290,
\href{http://arXiv.org/abs/gr-qc/0611043}{{\texttt{arXiv:gr-qc/0611043}}}.

\bibitem{Bohmer:2007wi}
C.~G. Boehmer and K.~Vandersloot, ``{Loop Quantum Dynamics of the Schwarzschild
  Interior},'' Phys. Rev. {\bf D76} (2007) 104030,
\href{http://arXiv.org/abs/0709.2129}{{\texttt{arXiv:0709.2129}}}.

\bibitem{Boehmer:2008fz}
C.~G. Boehmer and K.~Vandersloot, ``{Stability of the Schwarzschild Interior in
  Loop Quantum Gravity},'' Phys. Rev. {\bf D78} (2008) 067501,
\href{http://arXiv.org/abs/0807.3042}{{\texttt{arXiv:0807.3042}}}.

\bibitem{Modesto:2009ve}
L.~Modesto and I.~Premont-Schwarz, ``{Self-dual Black Holes in LQG: Theory and
  Phenomenology},'' Phys. Rev. {\bf D80} (2009) 064041,
\href{http://arXiv.org/abs/0905.3170}{{\texttt{arXiv:0905.3170}}}.

\bibitem{Chiou:2012pg}
D.-W. Chiou, W.-T. Ni, and A.~Tang, ``{Loop quantization of spherically
  symmetric midisuperspaces and loop quantum geometry of the maximally extended
  Schwarzschild spacetime},''
\href{http://arXiv.org/abs/1212.1265}{{\texttt{arXiv:1212.1265}}}.

    
    
    \bibitem{Tibrewala:2012xb} 
  R.~Tibrewala,
  ``{Spherically symmetric Einstein-Maxwell theory and loop quantum gravity corrections,}''
  Class.\ Quant.\ Grav.\  {\bf 29}, 235012 (2012)
    \href{http://arXiv.org/abs/1207.2585}{{\texttt{arXiv:1207.2585}}}.

  
      \bibitem{Gambini:2013ooa} 
  R.~Gambini and J.~Pullin,
  ``{Loop quantization of the Schwarzschild black hole,}''
  Phys.\ Rev.\ Lett.\  {\bf 110}, no. 21, 211301 (2013)
   \href{http://arXiv.org/abs/1302.5265}{{\texttt{arXiv:1302.5265}}}.

     
          \bibitem{Corichi:2015xia} 
  A.~Corichi and P.~Singh,
  ``{Loop quantization of the Schwarzschild interior revisited,}''
  Class.\ Quant.\ Grav.\  {\bf 33}, no. 5, 055006 (2016)
  \href{http://arXiv.org/abs/1506.08015}{{\texttt{arXiv:1506.08015}}}.
  


\bibitem{Olmedo:2017lvt} 
  J.~Olmedo, S.~Saini and P.~Singh,
  ``{From black holes to white holes: a quantum gravitational, symmetric bounce,}''
  Class.\ Quant.\ Grav.\  {\bf 34}, no. 22, 225011 (2017)
  \href{http://arXiv.org/abs/1707.07333}{{\texttt{arXiv:1707.07333}}}.
  
    \bibitem{Cortez:2017alh} 
  J.~Cortez, W.~Cuervo, H.~A.~Morales-Técotl and J.~C.~Ruelas,
  ``{Effective loop quantum geometry of Schwarzschild interior,}''
  Phys.\ Rev.\ D {\bf 95}, no. 6, 064041 (2017)
   \href{http://arXiv.org/abs/1704.03362}{{\texttt{arXiv:1704.03362}}}
  
  \bibitem{BenAchour:2018khr}  
  J.~Ben Achour, F.~Lamy, H.~Liu and K.~Noui,
  ``{Polymer Schwarzschild black hole: An effective metric,}''
  EPL {\bf 123}, no. 2, 20006 (2018)
     \href{http://arXiv.org/abs/1803.01152}{{\texttt{arXiv:1803.01152}}}.
  
  \bibitem{Bojowald:2018xxu} 
  M.~Bojowald, S.~Brahma and D.~h.~Yeom,
  ``{Effective line elements and black-hole models in canonical loop quantum gravity,}''
  Phys.\ Rev.\ D {\bf 98}, no. 4, 046015 (2018)
     \href{http://arXiv.org/abs/1803.01119}{{\texttt{arXiv:1803.01119}}}.

\bibitem{Ashtekar:2018cay} 
  A.~Ashtekar, J.~Olmedo and P.~Singh,
  ``{Quantum extension of the Kruskal spacetime,}''
  Phys.\ Rev.\ D {\bf 98}, no. 12, 126003 (2018)
  \href{http://arXiv.org/abs/1806.02406}{{\texttt{arXiv:1806.02406}}}.
  
      \bibitem{Bodendorfer:2019xbp} 
  N.~Bodendorfer, F.~M.~Mele and J.~Münch,
  ``{A note on the Hamiltonian as a polymerisation parameter,}''
  Class.\ Quant.\ Grav.\  {\bf 36}, no. 18, 187001 (2019)
    \href{http://arXiv.org/abs/1902.04032}{{\texttt{arXiv:1902.04032}}}.
    
        \bibitem{Bojowald:2019dry} 
  M.~Bojowald,
  ``{Comment (2) on "Quantum Transfiguration of Kruskal Black Holes",}''
  \href{http://arXiv.org/abs/1906.04650}{{\texttt{arXiv:1906.04650}}}.
  
  \bibitem{Bouhmadi-Lopez:2019hpp}   
  M.~Bouhmadi-Lopez, S.~Brahma, C.~Y.~Chen, P.~Chen and D.~h.~Yeom,
  ``{Comment on "Quantum Transfiguration of Kruskal Black Holes",}''
    \href{http://arXiv.org/abs/1902.07874}{{\texttt{arXiv:1902.07874}}}.
  
  \bibitem{Bodendorfer:2019cyv} 
  N.~Bodendorfer, F.~M.~Mele and J.~Münch,
  ``{Effective Quantum Extended Spacetime of Polymer Schwarzschild Black Hole,}''
  Class.\ Quant.\ Grav.\  {\bf 36}, no. 19, 195015 (2019)
  \href{http://arXiv.org/abs/1902.04542}{{\texttt{arXiv:1902.04542}}}.
  
  \bibitem{Bodendorfer:2019nvy} 
  N.~Bodendorfer, F.~M.~Mele and J.~Münch,
  ``{(b,v)-type variables for black to white hole transitions in effective loop quantum gravity,}''
    \href{http://arXiv.org/abs/1911.12646}{{\texttt{arXiv:1911.12646}}}.

  
  \bibitem{Bodendorfer:2019jay} 
  N.~Bodendorfer, F.~M.~Mele and J.~Münch,
  ``{Mass and Horizon Dirac Observables in Effective Models of Quantum Black-to-White Hole Transition,}''
    \href{http://arXiv.org/abs/1912.00774}{{\texttt{arXiv:1912.00774}}}.
    
    
      \bibitem{Bojowald:2015zha} 
  M.~Bojowald, S.~Brahma and J.~D.~Reyes,
  ``{Covariance in models of loop quantum gravity: Spherical symmetry,}''
  Phys.\ Rev.\ D {\bf 92}, no. 4, 045043 (2015)
    \href{http://arXiv.org/abs/1507.00329}{{\texttt{arXiv:1507.00329}}}.
    
    \bibitem{Aruga:2019dwq} 
  D.~Aruga, J.~Ben Achour and K.~Noui,
  ``{Deformed General Relativity and Quantum Black Holes Interior,}''
  \href{http://arXiv.org/abs/1912.02459}{{\texttt{arXiv:1912.02459}}}.
  
    
    \bibitem{BenAchour:2017ivq} 
  J.~Ben Achour, F.~Lamy, H.~Liu and K.~Noui,
  ``{Non-singular black holes and the Limiting Curvature Mechanism: A Hamiltonian perspective,}''
  JCAP {\bf 1805}, 072 (2018)
    \href{http://arXiv.org/abs/1712.03876}{{\texttt{arXiv:1712.03876}}}.
  
    \bibitem{Assanioussi:2019twp} 
  M.~Assanioussi, A.~Dapor and K.~Liegener,
  ``{Perspectives on the dynamics in a loop quantum gravity effective description of black hole interiors,}''
    \href{http://arXiv.org/abs/1908.05756}{{\texttt{arXiv:1908.05756}}}.
    
    \bibitem{Morales-Tecotl:2018ugi} 
  H.~A.~Morales-Tecotl, S.~Rastgoo and J.~C.~Ruelas,
  ``Effective dynamics of the Schwarzschild black hole interior with inverse triad corrections,''
      \href{http://arXiv.org/abs/1806.05795}{{\texttt{arXiv:1806.05795}}}.

    
      \bibitem{Alesci:2018loi} 
  E.~Alesci, S.~Bahrami and D.~Pranzetti,
  ``{Quantum evolution of black hole initial data sets: Foundations,}''
  Phys.\ Rev.\ D {\bf 98}, no. 4, 046014 (2018)
   \href{http://arXiv.org/abs/1807.07602}{{\texttt{arXiv:1807.07602}}}.
    
    \bibitem{Alesci:2019pbs} 
  E.~Alesci, S.~Bahrami and D.~Pranzetti,
  ``{Quantum gravity predictions for black hole interior geometry,}''
  Phys.\ Lett.\ B {\bf 797}, 134908 (2019)
      \href{http://arXiv.org/abs/1904.12412}{{\texttt{arXiv:1904.12412}}}.
      
      \bibitem{Bojowald:2005qw}
M.~Bojowald, R.~Goswami, R.~Maartens, and P.~Singh, ``{A Black hole mass
  threshold from non-singular quantum gravitational collapse},'' Phys. Rev.
  Lett. {\bf 95} (2005) 091302,
\href{http://arXiv.org/abs/gr-qc/0503041}{{\texttt{arXiv:gr-qc/0503041}}}.

\bibitem{Tavakoli:2013lga}
Y.~Tavakoli, J.~Marto, and A.~Dapor, ``{Dynamics of apparent horizons in
  quantum gravitational collapse},'' Springer Proc. Math. Stat. {\bf 60} (2014)
  427--431,
\href{http://arXiv.org/abs/1306.3458}{{\texttt{arXiv:1306.3458}}}.

\bibitem{Tavakoli:2013rna}
Y.~Tavakoli, J.~Marto, and A.~Dapor, ``{Semiclassical dynamics of horizons in
  spherically symmetric collapse},'' Int. J. Mod. Phys. {\bf D23} (2014),
  no.~7, 1450061,
\href{http://arXiv.org/abs/1303.6157}{{\texttt{arXiv:1303.6157}}}.

\bibitem{Bambi:2013caa}
C.~Bambi, D.~Malafarina, and L.~Modesto, ``{Non-singular quantum-inspired
  gravitational collapse},'' Phys. Rev. {\bf D88} (2013) 044009,
\href{http://arXiv.org/abs/1305.4790}{{\texttt{arXiv:1305.4790}}}.

\bibitem{Liu:2014kra}
Y.~Liu, D.~Malafarina, L.~Modesto, and C.~Bambi, ``{Singularity avoidance in
  quantum-inspired inhomogeneous dust collapse},'' Phys. Rev. {\bf D90} (2014),
  no.~4, 044040,
\href{http://arXiv.org/abs/1405.7249}{{\texttt{arXiv:1405.7249}}}.

\bibitem{These}
S.~Campbell, ``{Models of Non-Singular Gravitational collapse},'' Ph.D Thesis {\bf
  London SW7 2AZ}
(2014).
      
              \bibitem{Joe:2014tca} 
  A.~Joe and P.~Singh,
  ``{Kantowski-Sachs spacetime in loop quantum cosmology: bounds on expansion and shear scalars and the viability of quantization prescriptions,}''
  Class.\ Quant.\ Grav.\  {\bf 32}, no. 1, 015009 (2015)
     \href{http://arXiv.org/abs/1407.2428}{{\texttt{arXiv:1407.2428}}}
   
  \bibitem{Gambini:2014qga} 
  R.~Gambini and J.~Pullin,
  ``{Quantum shells in a quantum space-time,}''
  Class.\ Quant.\ Grav.\  {\bf 32}, no. 3, 035003 (2015)
     \href{http://arXiv.org/abs/1408.4635}{{\texttt{arXiv:1408.4635}}}
     
     \bibitem{Ziprick:2016ogy} 
  J.~Ziprick, J.~Gegenberg and G.~Kunstatter,
  ``{Polymer Quantization of a Self-Gravitating Thin Shell,}''
  Phys.\ Rev.\ D {\bf 94}, no. 10, 104076 (2016)
   \href{http://arXiv.org/abs/1609.06665}{{\texttt{arXiv:1609.06665}}}
     

  
  \bibitem{BenAchour:2016brs} 
  J.~Ben Achour, S.~Brahma and A.~Marciano,
  ``{Spherically symmetric sector of self dual Ashtekar gravity coupled to matter: Anomaly-free algebra of constraints with holonomy corrections,}''
  Phys.\ Rev.\ D {\bf 96}, no. 2, 026002 (2017)
  \ \href{http://arXiv.org/abs/1608.07314}{{\texttt{arXiv:1608.07314}}}
  
  \bibitem{BenAchour:2017jof} 
  J.~Ben Achour and S.~Brahma,
  ``{Covariance in self dual inhomogeneous models of effective quantum geometry: Spherical symmetry and Gowdy systems,}''
  Phys.\ Rev.\ D {\bf 97}, no. 12, 126003 (2018)
    \ \href{http://arXiv.org/abs/1712.03677}{{\texttt{arXiv:1712.03677}}}
  
  \bibitem{Bojowald:2019fkv} 
  M.~Bojowald, S.~Brahma, D.~Ding and M.~Ronco,
  ``{Deformed covariance in spherically symmetric vacuum models of loop quantum gravity: Consistency in Euclidean and self-dual gravity,}''
   \ \href{http://arXiv.org/abs/1910.10091}{{\texttt{arXiv:1910.10091}}}
  
  
 

 
\bibitem{Carballo-Rubio:2019nel} 
  R.~Carballo-Rubio, F.~Di Filippo, S.~Liberati and M.~Visser,
  ``{Opening the Pandora's box at the core of black holes,}''
  \href{http://arXiv.org/abs/1908.03261}{{\texttt{arXiv:1908.03261}}}.
  
  \bibitem{Carballo-Rubio:2019fnb} 
  R.~Carballo-Rubio, F.~Di Filippo, S.~Liberati and M.~Visser,
  ``{Geodesically complete black holes,}''
  \href{http://arXiv.org/abs/1911.11200}{{\texttt{arXiv:1911.11200}}}.
  
\bibitem{paperUS} 
  J.~Ben Achour, S.~Brahma and J.~P.~Uzan,
  ``{Bouncing compact objects I: Quantum extension of the Oppenheimer-Snyder collapse,}''
  JCAP {\bf 2003}, no. 03, 041 (2020)
  \href{http://arXiv.org/abs/2001.06148}{{\texttt{arXiv:2001.06148}}}.
  
  \bibitem{Senovilla:2013vra}
J.~M.M. Senovilla,
``{Junction conditions for F(R)-gravity and their consequences,}''
Phys. Rev. D \textbf{88}, 064015 (2013)
  \href{http://arXiv.org/abs/1303.1408}{{\texttt{arXiv:1303.1408}}}.
  
      \bibitem{D}
G.~Darmois, 
''{M\'emorial des Sciences Math\'ematiques}'', 
Fascicule 25 (Gauthier-Villars, Paris, 1927)

\bibitem{Israel:1966rt} 
W.~Israel,
``{Singular hypersurfaces and thin shells in general relativity,}''
Nuovo Cim. B \textbf{44S10}, 1 (1966)
\href{http://doi:10.1007/BF02710419}{{\texttt{doi:10.1007/BF02710419}}}.


\bibitem{Barrabes:1991ng}
C.~Barrabes and W.~Israel,
``{Thin shells in general relativity and cosmology: The Lightlike limit,}''
Phys. Rev. D \textbf{43}, 1129-1142 (1991)
\href{http://doi:10.1103/PhysRevD.43.1129}{{\texttt{doi:10.1103/PhysRevD.43.1129}}}.
 
 \bibitem{DeLorenzo:2014pta} 
  T.~De Lorenzo, C.~Pacilio, C.~Rovelli and S.~Speziale,
  ``{On the Effective Metric of a Planck Star,}''
  Gen.\ Rel.\ Grav.\  {\bf 47}, no. 4, 41 (2015)
    \href{http://arXiv.org/abs/1412.6015}{{\texttt{arXiv:1412.6015}}}.


\bibitem{Barcelo:2017lnx}
C.~Barcelo, R.~Carballo-Rubio, and L.~J. Garay, ``{Gravitational wave echoes
  from macroscopic quantum gravity effects},'' JHEP {\bf 05} (2017) 054,
\href{http://arXiv.org/abs/1701.09156}{{\texttt{arXiv:1701.09156}}}.

\bibitem{Carballo-Rubio:2018pmi}
R.~Carballo-Rubio, F.~Di~Filippo, S.~Liberati, C.~Pacilio, and M.~Visser, ``{On
  the viability of regular black holes},'' JHEP {\bf 07} (2018) 023,
\href{http://arXiv.org/abs/1805.02675}{{\texttt{arXiv:1805.02675}}}.

\bibitem{Carballo-Rubio:2018jzw}
R.~Carballo-Rubio, F.~Di~Filippo, S.~Liberati, and M.~Visser,
  ``{Phenomenological aspects of black holes beyond general relativity},''
  Phys. Rev. {\bf D98} (2018), no.~12, 124009,
\href{http://arXiv.org/abs/1809.08238}{{\texttt{arXiv:1809.08238}}}.

\bibitem{Brandenberger:2016vhg}
R.~Brandenberger and P.~Peter, ``{Bouncing Cosmologies: Progress and
  Problems},'' Found. Phys. {\bf 47} (2017), no.~6, 797--850,
\href{http://arXiv.org/abs/1603.05834}{{\texttt{arXiv:1603.05834}}}.

\bibitem{Battefeld:2014uga}
D.~Battefeld and P.~Peter, ``{A Critical Review of Classical Bouncing
  Cosmologies},'' Phys. Rept. {\bf 571} (2015) 1--66,
\href{http://arXiv.org/abs/1406.2790}{{\texttt{arXiv:1406.2790}}}.




\bibitem{Ashtekar:2006es}  
  A.~Ashtekar, T.~Pawlowski, P.~Singh and K.~Vandersloot,
  ``{Loop quantum cosmology of k=1 FRW models,}''
  Phys.\ Rev.\ D {\bf 75}, 024035 (2007)
    \href{http://arXiv.org/abs/0612104}{{\texttt{arXiv:0612104}}}.
    
    \bibitem{Szulc:2006ep} 
  L.~Szulc, W.~Kaminski and J.~Lewandowski,
  ``{Closed FRW model in Loop Quantum Cosmology,}''
  Class.\ Quant.\ Grav.\  {\bf 24}, 2621 (2007)
    \href{http://arXiv.org/abs/0612101}{{\texttt{arXiv:0612101}}}.

\bibitem{Corichi:2011pg} 
  A.~Corichi and A.~Karami,
  ``{Loop quantum cosmology of k=1 FRW: A tale of two bounces,}''
  Phys.\ Rev.\ D {\bf 84}, 044003 (2011)
  \href{http://arXiv.org/abs/1105.3724}{{\texttt{arXiv:1105.3724}}}.
  
  \bibitem{Dupuy:2016upu} 
  J.~L.~Dupuy and P.~Singh,
  ``{Implications of quantum ambiguities in $k$=1 loop quantum cosmology: distinct quantum turnarounds and the super-Planckian regime,}''
  Phys.\ Rev.\ D {\bf 95}, no. 2, 023510 (2017)
    \href{http://arXiv.org/abs/1608.07772}{{\texttt{arXiv:1608.07772}}}.
    
   \bibitem{Oppenheimer:1939ue}
J.~R. Oppenheimer and H.~Snyder, ``{On Continued gravitational contraction},''
  Phys. Rev. {\bf 56} (1939)
455--459.

 



\bibitem{Corichi:2013usa} 
  A.~Corichi and A.~Karami,
  ``{Loop quantum cosmology of k = 1 FLRW: Effects of inverse volume corrections,}''
  Class.\ Quant.\ Grav.\  {\bf 31}, 035008 (2014)
   \href{http://arXiv.org/abs/1307.7189}{{\texttt{arXiv:1307.7189}}}
   
   \bibitem{Dupuy:2019ibu} 
  J.~L.~Dupuy and P.~Singh,
  ``{Hysteresis and beats in loop quantum cosmology,}''
  \href{http://arXiv.org/abs/1912.11490}{{\texttt{arXiv:1912.11490}}}
  
  \bibitem{Singh:2013ava} 
  P.~Singh and E.~Wilson-Ewing,
  ``{Quantization ambiguities and bounds on geometric scalars in anisotropic loop quantum cosmology,}''
  Class.\ Quant.\ Grav.\  {\bf 31}, 035010 (2014)
    \href{http://arXiv.org/abs/1310.6728}{{\texttt{arXiv:1310.6728}}}
  
    \bibitem{Rovelli:2013zaa} 
  C.~Rovelli and E.~Wilson-Ewing,
  ``Why are the effective equations of loop quantum cosmology so accurate?,''
  Phys.\ Rev.\ D {\bf 90}, no. 2, 023538 (2014)
   \href{http://arXiv.org/abs/1310.8654}{{\texttt{arXiv:1310.8654}}}.
   
    
\bibitem{next}
J.~Ben Achour, S.~Brahma, S.~Mukohyama and J.~P.~Uzan,
``Towards consistent black-to-white hole bounces from matter collapse,''
JCAP \textbf{09}, 020 (2020)
 \href{http://arXiv.org/abs/2004.12977}{{\texttt{arXiv:2004.12977}}}.
 
 \bibitem{Kuchar:1994zk}
K.~V.~Kuchar,
``Geometrodynamics of Schwarzschild black holes,''
Phys. Rev. D \textbf{50}, 3961-3981 (1994)
 \href{http://arXiv.org/abs/9403003}{{\texttt{arXiv:9403003}}}.

   

\end{thebibliography}

\end{document}